\documentclass[11pt]{article}
\pdfoutput=1
\usepackage{dcolumn}
\usepackage{bm}
\usepackage{graphicx}
\usepackage{amssymb,amsmath}
\usepackage{multirow}
\usepackage{afterpage}
\usepackage{cite,color,url}
\usepackage[colorlinks=true
,urlcolor=blue
,anchorcolor=blue
,citecolor=blue
,filecolor=blue
,linkcolor=blue
,menucolor=blue
,linktocpage=true
,pdfproducer=medialab
,pdfa=true
]{hyperref}

\usepackage{array,booktabs}
\usepackage{subcaption}

\usepackage{slashed}
\usepackage{epsfig,psfrag,rotating,soul}
\usepackage{rotfloat}
\usepackage[font={small}]{caption}
\usepackage{colortbl}

\evensidemargin \oddsidemargin
\allowdisplaybreaks

\let\OLDthebibliography\thebibliography
\renewcommand\thebibliography[1]{
  \OLDthebibliography{#1}
  \setlength{\parskip}{0pt}
  \setlength{\itemsep}{0pt plus 0.3ex}
}

\definecolor{mygray}{gray}{0.85} 
\definecolor{myblue}{cmyk}{0.65, 0.37, 0.0, 0.19}

\begin{document}
\thispagestyle{empty}

\def\thefootnote{\fnsymbol{footnote}}

\vspace*{1cm}

\begin{center}

\begin{Large}
\textbf{\textsc{The Elusive Muonic WIMP }}
\end{Large}

\vspace{1cm}

{\sc
Anibal~D.~Medina$^{1}$%
\footnote{{\tt \href{mailto:anibal.medina@fisica.unlp.edu.ar}{anibal.medina@fisica.unlp.edu.ar}}}%
, Nicol\'as~I.~Mileo$^{1}$%
\footnote{{\tt \href{mailto:mileo@fisica.unlp.edu.ar}{mileo@fisica.unlp.edu.ar}}}%
, Alejandro Szynkman$^{1}$%
\footnote{{\tt \href{mailto:szynkman@fisica.unlp.edu.ar}{szynkman@fisica.unlp.edu.ar}}}%
and\\ Santiago A. Tanco$^{1}$%
\footnote{{\tt \href{mailto:santiago.tanco@fisica.unlp.edu.ar}{santiago.tanco@fisica.unlp.edu.ar}}}%
}

\vspace*{.7cm}

{\sl

$^1$IFLP, CONICET - Departamento de F\'{\i}sica, Universidad Nacional de La Plata, \\ 
C.C. 67, 1900 La Plata, Argentina

}

\end{center}

\vspace{0.1cm}

\begin{abstract}
\noindent
The Weakly Interacting Massive Particle (WIMP) paradigm is one of the most popular scenarios for Dark Matter (DM) theories that however is strongly constrained, in particular by direct detection experiments. We stick with the WIMP hypothesis and consider a Dirac fermion candidate for DM that interacts with the Standard Model (SM) via a spin-1 $Z'$, arising from the spontaneous breaking of an Abelian $U(1)'_{\mu}$ gauge symmetry, under which only second generation leptons and the DM are appropriately charged. Due to the charge assignment, the model is gauge anomalous and can only be interpreted as an effective field theory (EFT) at low energy. The $Z'$ couples at tree level only to the vector DM current, to the axial muon current and to left-handed muonic neutrinos, so the WIMP-nucleon cross section is beyond the experimental reach of spin-independent (SI) direct detection searches. We focus on $Z'$ masses between 200 GeV and 500 GeV, and study the current bounds on the model coming from direct and indirect detection of DM, collider searches, contributions to $(g-2)_{\mu}$ and to neutrino trident production. We find that large regions of the parameter space remains to be explored. In the context of LHC searches, we study the impact of a muon-exclusive signal region for the $3\mu$ + ${E}^{\rm miss}_T$ channel with an invariant mass window around $m_{Z'}$. We show that this search can significantly improve the current collider bounds. Finally, from the anomalous nature of our EFT, there remain at low energy triboson anomalous interactions between the $Z'$ and the electroweak (EW) SM gauge bosons. We explore the possibilities of probing these interactions at the LHC and at a 100 TeV proton collider finding it extremely challenging. On the other hand, for a muon collider the inclusive resonant channel $\mu^{+}\mu^{-}\to Z'\to ZZ$ could be probed in the most promising scenario with a luminosity of $\mathcal{O}({\rm few}\; 10)$ ${\rm fb}^{-1}$.
\end{abstract}

\def\thefootnote{\arabic{footnote}}
\setcounter{page}{0}
\setcounter{footnote}{0}

\newpage

\section{Introduction}
\label{intro}

There exists an overwhelming gravitational evidence for the existence of a new type of matter which is stable on cosmological scales and neutral under Electromagnetism. Its origin and composition cannot be explained by the Standard Model (SM) of particle physics, so it represents one of the greatest puzzles of modern physics. Many theoretical ideas beyond the SM attempt to explain the nature of this dark matter (DM), and within the attempts to describe DM as a particle, one of the most popular ones is that of a weakly interacting massive particle (WIMP) that introduces a new particle associated with the DM with a mass and interactions to SM states that are of order as those expected for an electroweak (EW) state. In that case one naturally obtains annihilation cross sections for the DM that provide via the mechanism known as freeze-out a relic density $\Omega_{\mathrm{DM}}h^2\sim 0.1$, in accordance with measurements done by several astrophysical probes. 

Despite the great appealing of the WIMP paradigm, the physics community would have naturally expected its existence to be detected already by now. There are several searches that probe the non-gravitational interactions of the DM with the SM states. Among them, direct detection experiments that involve the scattering of DM particles against heavy nuclei impose some of the most restrictive constraints~\cite{DiGangi:2018dek,XENON100:2013ele,LUX:2016ggv}, in particular in the parameter space regions in which WIMPS would tend to naturally lie. In fact typical cross sections expected for WIMP scattering against nuclei are already ruled out by several orders of magnitude. As direct detection searches continue probing smaller scattering cross sections and no sign of DM is detected in them, the WIMP paradigm becomes further in tension with data although it still remains as a viable possibility~\cite{Arcadi:2017kky,GAMBIT:2021rlp,Carena:2019pwq,Blanco:2019hah}. Another probe of the non-gravitational interactions of the DM are what are known as indirect detection experiments that measure the signals of DM annihilation in high density regions of the Universe, such as the center of the Milky Way~\cite{Fermi-LAT:2015sau} or in DM dominated galaxies such as dwarf spheroidals (dSph)~\cite{Fermi-LAT:2013sme,Fermi-LAT:2015att}. Finally  it may also be possible to directly produce the DM in particle accelerators, which would then be indirectly detected as missing energy~\cite{ATLAS:2016bek,CMS:2017jdm,ATLAS:2017txd,ATLAS:2017bfj,CMS:2014jvv}.

In regards to direct detection experiments and the WIMP paradigm, given that the former test DM-nucleon cross sections, one may think that one way to avoid them is to propose a model in which DM is coupled to a leptophilic mediator~\cite{Fox:2008kb}, leading to vanishing scattering tree-level diagrams with nuclei. Direct detection searches have become however so constraining in the allowed scattering cross sections that even interactions with nuclei at 1-loop order (with leptons running in the loop) are under pressure~\cite{Haisch:2013uaa,Kopp:2009et,DEramo:2014nmf,DEramo:2017zqw}. In this work we propose to take the WIMP paradigm all the way to the end and so we consider a simple effective WIMP model that provides such small contribution to the DM scattering cross sections probed by direct detection experiments that they would naturally be buried under what is known as the neutrino scattering floor. We show that under simple assumptions about the nature of the interactions of the DM and its mediator to SM particles but sticking to masses and couplings of order EW, a model emerges in which second generation leptons and a Dirac fermion DM are charged vector-axially and vectorially respectively under a new $U(1)^{'}_{\mu}$ spontaneously broken gauge symmetry. Under these conditions, the WIMP miracle is still accomplished and at the same time we are able to avoid any signal that could be measured by direct detection experiments at present and in the future (all the way up to the neutrino scattering floor). We find that indirect detection signals from current searches done by the Fermi-LAT collaboration from signals at the center of our own galaxy and from dwarf spheroidal galaxies do not currently provide strong constraints on the model. There can be however interesting collider searches that probe certain regions of the parameter space from current and future measurements at the LHC. There are also potential contributions to neutrino trident production and the magnetic dipole moment of the muon that though they are unrelated to the DM interactions, provide strong constraints to the mediator mass and coupling strength to muons and muonic neutrinos. Furthermore, our effective model in particular being gauge anomalous provides a window into the UV physics that is responsible for curing the anomalies in the form of triple gauge boson couplings, as it has been analyzed in theories with an anomalous $U(1)^{'}$ within different contexts~\cite{Dror:2017nsg,Ismail:2017ulg,Ismail:2017fgq}. By studying these anomalous interactions, we show that though it appears to be hopeless probe them at the high luminosity LHC (even hard at a 100 TeV hadron collider), they may be able to be probed at a future muon collider providing a clear signal and a great opportunity as a window into the UV physics.
 
The paper is organized as follows. In Section~\ref{sec:model} we introduce the effective $U(1)'_{\mu}$ DM model describing its interactions. In Section~\ref{sec:expcon} we discuss the experimental constraints on the model coming from DM relic density calculations, direct and indirect detection DM experiments, collider searches as well as contributions to $(g-2)_{\mu}$ and neutrino trident production. In Section~\ref{Results}, we focus on the projections at the 14 TeV LHC with luminosities of 300 $\mathrm{fb}^{-1}$ and 3000 $\mathrm{fb}^{-1}$ for the $3\mu + E_T^{\mathrm{miss}}$ channel and show that large regions of parameter space can be probed if a  window for the invariant mass of the muon pair around the $Z'$ mass is added, though there are still regions in which the model remains elusive. Finally in Section~\ref{sec:anomalies}, we discuss a very interesting feature of our model which is the presence of anomaly induced triple gauge boson couplings and their possible collider signatures at the LHC, and at a hypothetical muon collider or a $100\;\mathrm{TeV}$ hadron collider. Our conclusions are given in Section~\ref{conclu}.

\section{The $U(1)'_\mu$ axial model}
\label{sec:model}

We decide to pursue the idea of obtaining a model in which the WIMP miracle happens in a simple way, for couplings and masses of order the EW scale (or not too far from it). The lack of signal in direct detection experiments leads us to consider a leptophilic DM model with a massive vector mediator between the visible and dark sector, that couples vector-axially (VA) to leptons and vectorially (V) to the DM. These kind of interactions (VA with leptons and V with DM) have been shown to provide vanishing contributions at all loop-levels to the mixing between the $Z'$ and photons which tend to dominate spin independent DM-nuclei scattering~\cite{Kopp:2009et,DEramo:2014nmf,DEramo:2017zqw} and only provide contributions to spin independent DM scattering via mass mixing between the $Z'$ and the $Z$ gauge boson, proportional to the lepton's Yukawa couplings, which would put interaction with third generation leptons in tension. There are also  $e^{+}e^{-}\to e^{+} e^{-}$ processes known as compositeness bounds at the Large Electron-Positron (LEP) collider that strongly constrain interactions with first generation leptons implying $m_{Z'}\gtrsim 3$ TeV in that case~\cite{DEramo:2017zqw}. Thus we decide to consider vector-axial interactions only with second generation leptons and vector interactions with the DM, which furthermore we choose to be Dirac such that in the annihilation process the s-wave contribution in the non-relativistic limit is non-vanishing.

\begin{figure}[h]
\centering
\includegraphics[scale=0.4]{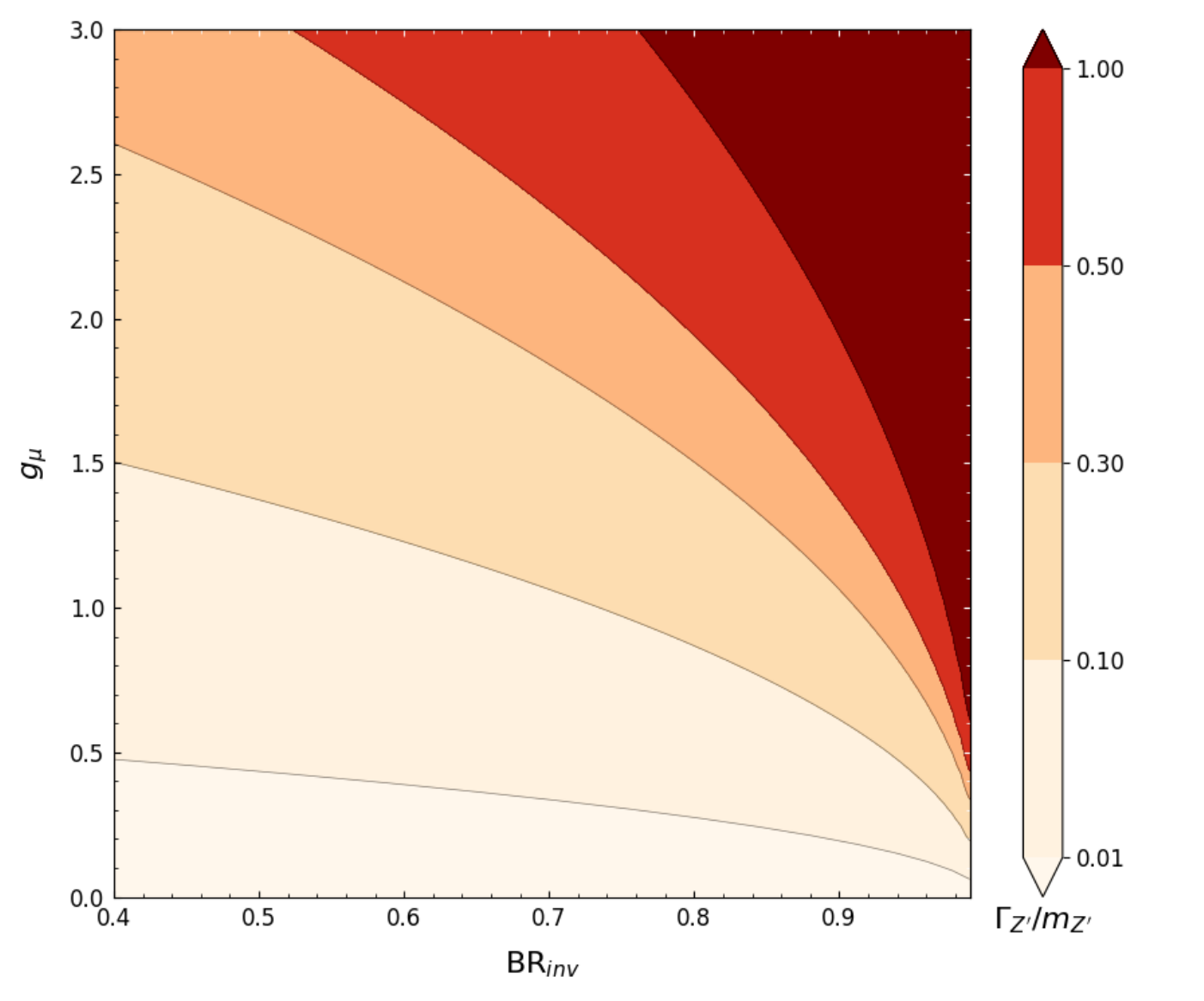}
\caption{Values of $\Gamma/m_{Z'}$ in the plane $\mathrm{BR}_{\mathrm{inv}}$-$g_{\mu}$ according to Eq.~(\ref{eq6}). In our analysis, we restrict ourselves to the region $\Gamma_{Z'}/m_{Z'}<0.3$.}
\label{figGtoM}
\end{figure}

Specifically, we introduce a new Abelian gauge symmetry $U(1)'_\mu$ under which second generation leptons are charged, such that
the interaction in the mass basis for muons are axial and we also introduce a Dirac fermion $\chi$ charged vectorially under  $U(1)'_\mu$, which we identify with the DM; the rest of the SM fields including the Standard Model Higgs remain neutral under $U(1)'_\mu$~\footnote{A model with a muon-philic $Z'$ coupled to vector-like leptons has been recently studied in~\cite{Kawamura:2019rth}. For muon-philic models in relation to dark matter see~\cite{Abdughani:2021oit,Perelstein:2020suc}}. This new gauge symmetry is assumed to be spontaneously broken at some larger scale (possibly {\it \`a la Higgs}, via the vacuum expectation value of a scalar field) and we work in the effective theory of the SM matter and gauge field  content, with the addition of the Dirac DM $\chi$ and a massive vector gauge $Z'$, \footnote{In the following, we consider the Higgs field responsible for the spontaneous breaking of the $U(1)'_\mu$ to be heavy enough to be integrated out from the EFT. Moreover, the $Z'$ mass can also arise from the Stückelberg mechanism, and a new Higgs would not be necessary.}
\begin{equation}
\label{eq1}
    \mathcal{L}=\mathcal{L}_{SM} + \bar{\chi}(i\gamma^\rho\partial_\rho-m_\chi)\chi-\frac{1}{4}Z'_{\mu\nu}Z'^{\mu\nu} + \frac{1}{2}m_{Z'} Z'_\rho Z'^\rho + g_\chi \bar{\chi}\gamma^\rho\chi Z'_\rho + g_\mu \bar{\mu}\gamma^\rho\gamma^5\mu Z'_\rho - g_\mu \bar{\nu}_{\mu L}\gamma^\rho\nu_{\mu L} Z'_\rho \; , \vspace*{0.5cm}
\end{equation}
where $Z'_{\mu\nu}=\partial_{\mu}Z'_{\nu}-\partial_{\nu}Z'_{\mu}$ is the $Z'$-field strength,  $g_\chi=Q_{\chi} g'$, $g_\mu=Q_{\mu}g'$ are the coupling strengths for the DM and charged muon interactions with respective charges $Q_{\chi}$ and $Q_{\mu}$ under $U(1)'_\mu$, $g'$ the $U(1)'_\mu$ coupling, and $m_\chi$ and $m_{Z'}$ are the masses of the DM and the $Z'$ gauge boson. The $Z'-\nu_\mu$ coupling is fixed to preserve the EW symmetry. We assume a vanishing tree-level kinetic mixing term between the $Z'$ and the SM EW gauge bosons.
In addition, we work in the scenario with $m_{Z'}>2m_{\chi}$ in which the phenomenology at the LHC is enriched. When $m_{Z'}<2m_{\chi}$, the only invisible decay channel is $Z'\to \nu_{\mu}\bar{\nu}_{\mu}$ so that the invisible branching ratio is $1/3$ irrespective of the values of $m_{\chi}$ or $g_{\chi}$ and the phenomenology of the LHC searches is disconnected from the DM parameters of the model.
From the interactions in Eq.~(\ref{eq1}) we obtain the following expressions for the partial decay widths of the $Z'$ boson at leading order:
\begin{eqnarray}
\label{eq2}
\Gamma(Z'\to \mu^+\mu^-)&=& \frac{g^2_{\mu}m_{Z'}}{12\pi}(1-4z_{\mu})^{3/2}\\[1mm]
\label{eq3}
\Gamma(Z'\to \nu_{\mu}\bar{\nu}_{\mu})&=& \frac{g^2_{\mu}m_{Z'}}{24\pi}\\[1mm]
\label{eq4}
\Gamma(Z'\to \chi\bar{\chi})&=& \frac{g^2_{\chi}m_{Z'}}{12\pi}\sqrt{1-4z_{\chi}}(1+2z_{\chi}),
\end{eqnarray}
where $z_{x}=m^2_x/m^2_{Z'}$. If the total width is saturated by the channels in Eqs.~(\ref{eq2})-(\ref{eq4}), then its  branching ratio can be obtained as $\mathrm{BR}_{Z'}=\mathrm{BR}(Z'\to \mu^+\mu^-)+\mathrm{BR}_{\mathrm{inv}}$, where $\mathrm{BR}_{\mathrm{inv}}\equiv\mathrm{BR}(Z'\to\nu_{\mu}\bar{\nu}_{\mu})+\mathrm{BR}(Z'\to \chi\bar{\chi})$ is the branching ratio into states that appear at the LHC as missing transverse energy and can be written as,
\begin{equation}
\label{eq5}
\mathrm{BR}_{\mathrm{inv}}= \frac{1+2\xi^2(1-4z_{\chi})^{1/2}(1+2z_{\chi})}{3+2\xi^2(1-4z_{\chi})^{1/2}(1+2z_{\chi})},
\end{equation} 
where we have used the definition $\xi=g_{\chi}/g_{\mu}$. From Eqs.~(\ref{eq2})-(\ref{eq5}), the ratio between the total decay width, $\Gamma_{Z'}$, and the $Z'$ mass, can be parametrized in terms of the coupling $g_{\mu}$ and $\mathrm{BR}_{\mathrm{inv}}$ as follows,
\begin{equation}
\label{eq6}
\frac{\Gamma_{Z'}}{m_{Z'}}= g^2_{\mu}\left(\frac{1}{8\pi}+ \frac{3\,\mathrm{BR}_{\mathrm{inv}}-1}{24\pi(1-\mathrm{BR}_{\mathrm{inv}})}\right).
\end{equation}
In the analysis of the model we will restrict ourselves to scenarios in which $\Gamma_{Z'}/m_{Z'}<0.3$. It follows then that this expression is useful to translate this bound, for each value of $\mathrm{BR}_{\mathrm{inv}}$, into a bound for the coupling $g_{\mu}$. The requirement $\Gamma_{Z'}/m_{Z'}<0.3$ allows to have a not too wide $Z'$ while keeping available regions of the parameter space that are interesting from a phenomenological standpoint \footnote{Notice that in our analysis we will not rely on the narrow width approximation.}. For example, as can be seen from Fig.~\ref{figGtoM}, for $\mathrm{BR}_{\mathrm{inv}}$ as large as 0.9 we can still probe $g_{\mu}$ values up to $\sim 1$.

Given that in particular the electromagnetically charged muon is also charged under the new $U(1)'_{\mu}$ gauge symmetry, one cannot write the usual Yukawa interaction for the muon responsible via EW symmetry breaking to generate the muon mass since it would explicitly break $U(1)'_{\mu}$. This interaction can however arise from a higher-dimensional operator involving the usual Yukawa interaction combined with a SM singlet Higgs field responsible for the spontaneous breaking of the $U(1)'_{\mu}$ gauge symmetry, which we shall denote $\Phi$, with charge $Q_{\Phi}$ with respect to $U(1)'_{\mu}$ and whose vacuum expectation value (vev) we denote $\langle \Phi\rangle\equiv f\gg v_{EW}=246$ GeV sets the scale at which $U(1)'_{\mu}$ is broken~\footnote{For the $Z'$ gauge boson to remain in the low energy effective theory, $|Q_{\Phi}|<1$, since for an Abelian group that is spontaneously broken, $m_{Z'}\sim |Q_{\Phi}| f$. Also notice that in order for Eq.~(\ref{HighOp}) to be truly a higher-order operator we must demand that $Q_{\Phi}<0$ if $Q_{\mu}>0$.},
\begin{equation}
\mathcal{L}_{\rm Yuk}=-\lambda_{\mu}\left(\frac{\Phi}{M}\right)^{-2\frac{Q_{\mu}}{Q_{\Phi}}}\bar{L}H\mu_R+h.c,\label{HighOp}
\end{equation}
where $\lambda_{\mu}$ is a dimensionless coupling of order one and $M\gg f$ is the scale of the new physics that generates this operator. It is then clear that once $\Phi$ acquires a vev the low energy muon Yukawa takes the value $y_{\mu}=\lambda_{\mu}(f/M)^{-2Q_{\mu}/Q_{\Phi}}$, that can naturally accommodate the measured muon mass due to its smallness. We assume that this or a similar mechanism is responsible for the generation of the muon Yukawa and simply work with the low energy muonic Yukawa interaction included in $\mathcal{L}_{SM}$ in Eq.~(\ref{eq1}).

Notice also that the fact that we have chosen a VA gauge interaction for which only the second generation leptons are charged implies naively the appearance of gauge anomalies which are  known to lead to inconsistencies at the quantum level either by the breaking of unitarity and/or gauge invariance. Of course, the quantum field theory which completes our effective theory  cannot have these anomalies in the UV and therefore there must exist additional fermions which cancel the possible gauge anomalies present. The effect of these additional fermions in the low energy effective theory manifest itself via the presence of triple gauge bosons couplings involving $U(1)'_{\mu}$ and the SM gauge bosons. We shall comment more on this in Section~\ref{sec:anomalies} and its possible phenomenological collider signatures in Sections~\ref{sec:anomcolliderhadron} and~\ref{sec:mucoll}.

\section{Experimental constraints}
\label{sec:expcon}
In this section we study the conditions that must be met in order to reproduce the correct relic density of DM in the Universe today. Furthermore, we look into the possible DM standard signatures in direct and indirect detection experiments, as well as more specific signals that are particular of our model such as neutrino trident production and the muon anomalous magnetic moment. These latter affect indirectly the DM predictions by constraining the $Z'$ interaction with the second generation leptons. We also comment on the current collider constraints in the form of final states with muons and missing energy.

\subsection{Relic density}
\label{sec:relic}
The first main prediction we should satisfy is the DM relic abundance as measured by the Planck collaboration, $\Omega_{\mathrm{DM}}h^2=\rho_{\mathrm{DM}}/\rho_c=0.120\pm 0.001$~\cite{Planck:2018vyg}, where $\rho_{\mathrm{DM}}$ is the DM density and $\rho_c$ the critical density of the Universe.  As it is well-known, the WIMP miracle occurs within the standard \textit{freeze out} mechanism, where initial conditions are washed out once the DM candidate enters in thermal equilibrium with the early Universe plasma. Given that the coupling strengths are assumed to be of order EW but not much smaller, it is clear that at earlier times in the Universe, the DM candidate $\chi$ can easily reach equilibrium with the thermal bath at temperature $T$ via the process,
\begin{equation}
    \chi \bar{\chi} \longleftrightarrow \ell \bar{\ell} \; ,
\end{equation}
where $\ell = \mu, \nu_\mu$. The DM density then follows the thermal equilibrium density and as the temperature in the Universe drops due to the expansion it ends up following a Maxwell-Boltzmann distribution which leads to an exponential decrease in the DM density once the energy available is not sufficient to produce DM pairs from the thermal bath.  However, once the annihilation rate is smaller than the Hubble expansion of the Universe at some temperature $T_F\approx m_{\chi}/25$,  the DM particles fall out of  equilibrium and decouple from the thermal bath, \textit{freezing} at a density that gets diluted only by the expansion of the Universe. We will not demand that $\chi$ makes up all of the DM (though we will certainly consider this case as well), but instead we  demand that it does not lead to a closed Universe by requiring the following condition to hold,
\begin{equation}
   \Omega_{\chi} h^2 \leq \Omega_{\mathrm{DM}} h^2=0.120 \; .
\end{equation}
which allows more freedom for our parameter space by assuming that some other source completes the full DM content of the Universe. All of this is done by integrating Boltzmann equation numerically and computing $\Omega_{\chi} h^2$ for a particular point in our parameter space. We perform this automatically by using the publicly available {\tt micrOMEGAs} code~\cite{Belanger:2010pz}.

\subsection{Direct detection}

Direct detection experiments aim to detect DM particles present in the gravitationally bound halo of our galaxy. Through scattering with subatomic particles, DM could leave a signal in underground detectors, in the form of recoil energy. The most stringent bounds on DM-nucleon scattering cross sections come from these experiments, in particular for spin independent (SI) interactions for which the scattering cross sections are enhanced by a coherent sum of the nucleons in the nucleus. This leads to strong constraints for the possible effective interactions of WIMPS with quarks as shown in~\cite{ParticleDataGroup:2020ssz,DiGangi:2018dek,XENON100:2013ele,LUX:2016ggv}. In order to avoid these bounds we have chosen as mentioned before a leptophilic $Z'$ mediator with axial couplings to muons, which guarantees that not only tree level diagrams are not present, but also that $Z'$-photon mixing vanishes at all loop orders~\footnote{The largest contributions to the scattering cross section comes from the $Z'$-photon mixing.}. There remains however an effective operator  that provides a contribution to both spin independent and spin dependent (SD) interactions and involves $Z$-$Z'$ mass mixing~\cite{DEramo:2017zqw}~\footnote{ 
Loop-induced $Z-Z'$ kinetic mixing is also possible, but it is suppressed with respect to mass mixing by $(q/M_Z)^2/y_\mu^2 \sim ({\rm MeV}/100\; {\rm GeV})^2/y_\mu^2 \sim \mathcal{O}(10^{-4})$, where $q$ is the momentum transfer in the scattering \cite{Kopp:2009et}.},
\begin{equation}
    \mathcal{L}_H = i g_H H^\dagger \overleftrightarrow{D}_\mu H Z'_\mu \; ,
\end{equation}
where $g_H$ is a loop-induced effective coupling proportional to the Yukawa coupling of the leptons that run in the loop and are charged under $U(1)'_{\mu}$, so its contribution is naturally suppressed (however, notice that a model with the tau charged under the $U(1)'$ symmetry is significantly constrained~\cite{DEramo:2017zqw}). One can calculate the corresponding quark-DM contact interactions that provide contributions to SI and SD interactions respectively,
\begin{equation}
    \mathcal{L} = \bar{\chi}\gamma_\mu\chi\sum_q\left[ C_V^{(q)}\bar{q}\gamma^\mu q +  C_A^{(q)}\bar{q}\gamma^\mu \gamma^5 q \right] \; , \label{DDcont}
\end{equation}
where the vector and axial coefficients $C_V^{(q)}$ and $C_A^{(q)}$ can be obtained from the renormalization group equations. We apply to our model the results given in~\cite{DEramo:2017zqw}, where the running of the coupling constants is performed from the  energy cutoff of the effective leptophilic model, $\Lambda_{UV}\sim f$, down to the energy scale $\mu_N$ associated with nuclear processes. These coefficients read,
\begin{align}
    C_{V}^{\left(q\right)}&=\frac{g_{\chi}g_{\mu}}{m_{Z'}^{2}}\frac{y_{\mu}^{2}}{4\pi^{2}}\left(T_{q}^{\left(3\right)}-2s_{W}^{2}Q_{q}\right)\mathrm{log}\left(\frac{\Lambda_{UV}}{\mu_{N}}\right)\; \label{CVq},\\C_{A}^{\left(q\right)}&=-\frac{g_{\chi}g_{\mu}}{m_{Z'}^{2}}\frac{y_{\mu}^{2}}{4\pi^{2}}T_{q}^{\left(3\right)}\mathrm{log}\left(\frac{\Lambda_{UV}}{\mu_{N}}\right) \label{CAq}\; ,
\end{align}
where $y_\mu$ is the Yukawa coupling for the muon, $T_q^{(3)}$ and $Q_q$ are the weak isospin and charge of the quark $q$ respectively, and $s_W$ is the sine of the Weinberg angle $\theta_W$. 

Only the first contact term of Eq.~\eqref{DDcont} contributes to a spin-independent signal, and thus we focus on this term only. The amplitude for the scattering of $\chi$ by a nucleon $N$ is then calculated as,
\begin{equation}
    \mathcal{M}=\langle \chi N | \sum_q C_V^{(q)} \bar{\chi} \gamma_\mu\chi \bar{q}\gamma^\mu q |\chi N\rangle \; .
\end{equation}
When evaluating matrix elements of quarks in a nucleon $N$ at zero momentum transfer, only valence quarks amount for the result~\cite{Lin:2019uvt},
\begin{equation}
    \langle N | \bar{q}\gamma^\mu q |N \rangle = \bar{u}_N(k') \left( F_1^{q}(q^2)\gamma^\mu + \frac{i}{2m_N} F_2^{q}(q^2)\sigma^{\mu\nu}q_\nu\right) u_N(k) \; \stackrel{q^{2}\rightarrow 0}{\sim}\;n_q\bar{u}_N\gamma^\mu u_N \; ,
\end{equation}
where the first form factor is approximated by the number of valence quarks $n_q$ in the nucleon, and the second term is dropped in the $q^2=0$ approximation. Also, in the non-relativistic regime we can write for both nucleon and DM parts, 
\begin{equation}
    \bar{u}\gamma^\mu u \sim \bar{u}\gamma^0 u \sim 2m \xi_{s'}^\dagger\xi_s\; ,
\end{equation}
where $\xi$ is a two-component spinor and $s$ and $s'$ are the initial and final spin states. Squaring the amplitude, averaging over initial states, summing over final states and integrating in the phase space we obtain the total cross section,
\begin{equation}
    \sigma_N = \frac{1}{\pi} \frac{(m_N m_\chi)^2}{(m_N + m_\chi)^2} \left( n_u C_V^{(u)} + n_d C_V^{(d)} \right)^2\;,
\end{equation}
and by setting $n_u=2$, $n_d=1$ we can obtain the $\chi$-proton cross section. Finally, we evaluate the $C_V^{(q)}$ coefficients at the benchmark values $g_\mu=g_\chi=1$, $m_{Z'}=200\;{\rm GeV}$. The nuclear scale is set at $\mu_N=1\;{\rm GeV}$ where the hadronic matrix elements are evaluated, and since we expect heavier states to appear at few TeV, we set for practical purposes $\Lambda_{UV}=10\;{\rm TeV}$, though the result only depends mildly on the scale $\Lambda_{UV}$. For the range of $\chi$ masses evaluated in this work, this results in a $\chi$-proton cross section of,
\begin{equation}
    \sigma_p \approx 1\times 10^{-54}\;{\rm cm^2} \;,
\end{equation}
and thus it is unconstrained by current experimental research, whose more restrictive upper bounds are around $10^{-47}\;{\rm cm^2}$~\cite{Aprile_2018}. Moreover, this cross section value sits well below the neutrino floor $\sim 10^{-50}\;{\rm cm^2}$, inside a region of the parameter space which seems to be inaccessible by conventional SI direct detection experiments~\footnote{Directional DM direct detection experiments may be able to distinguish coherent neutrino-nucleus scattering from DM-nucleus scattering~\cite{Mayet:2016zxu}. Probes below the neutrino floor could also be available with the detection of gravitational waves produced during the electroweak phase transition~\cite{2020}.}.

\subsection{Indirect detection}
\label{sec:indirect}
In our Universe today, dark matter annihilation is heavily suppressed by its low abundance. Thus it can only be observed in high density regions, such as galactic centers or DM-dominated dwarf galaxies. For our case, muon pairs are produced from $\chi\bar{\chi}$ annihilation, which then lead to a continuum spectrum in the form of  bremsstrahlung radiation, whose differential flux per unit energy per unit solid angle is given by,
\begin{equation}
    \frac{d^{2}\Phi\left(E,\mathbf{n}\right)}{dE\,d\Omega}=\frac{\left\langle \sigma v\right\rangle _{\chi}}{8\pi m_{\chi}^{2}}\frac{dN}{dE}\int d\ell\,\rho_{\ensuremath{\chi}}^{2}\left(\ell;\mathbf{n}\right)\; ,
\end{equation}
where $\left\langle \sigma v \right\rangle _{\chi}$ is the thermally-averaged $\chi$-annihilation cross section into muons, $dN/dE$ is the radiation spectrum generated by the muons, and  $\rho_\chi$ is the $\chi$-density which must be integrated along the line of sight. 

The Fermi-LAT collaboration is able to measure the $\gamma$-ray sky via its satellite experiment with unprecedented precision. In particular, it provides strong constraints on possible DM annihilation coming from the center of our Milky Way~\cite{Fermi-LAT:2015sau} and also from a series of dwarf spheroidal galaxies~\cite{Fermi-LAT:2013sme,Fermi-LAT:2015att}, which are expected to be DM dominated. Indirect detection experiments typically give constraints in the $\left(m_\chi,\left\langle \sigma v \right\rangle _{\chi}\right)$ plane, assuming $\chi$ accounts for all the DM in the Universe. Since as mentioned before we relax such assumption and only demand that the relic density should not overclose the Universe,  in order to translate the constraints from the Fermi-LAT results to our particular model, we rescale  $\rho_{\chi}$ as~\cite{Wu:2017iji},
\begin{equation}
  \rho_\chi = \frac{\Omega_{\chi}h^2}{\Omega_{\mathrm{DM}}h^2} \rho_{\mathrm{DM}}\; ,
\end{equation}
where $\Omega_{\mathrm{DM}}h^2=0.120$ and $\rho_{\mathrm{DM}}$ is the total DM density used in the indirect detection analysis. If, for a given $m_{\chi}$, the cross section is bounded above by $\left\langle \sigma v \right\rangle _{\mathrm{lim}}$, then this corresponds to the following limit in our model,
\begin{equation}
    \left(\frac{\Omega_{\chi}h^2}{\Omega_{\mathrm{DM}}h^2}\right)^2 \left\langle \sigma v \right\rangle _{\chi} \leq \left\langle \sigma v \right\rangle _{\mathrm{lim}} \; .
    \label{limReescID}
\end{equation}
We use the values for $\langle \sigma v \rangle_{\rm lim}$ as given in Fig.~10 of~\cite{Hoof_2020} at $\rm 95\%$ Confidence Level (CL), taking the more restrictive Bayesian limit. The relic abundance and cross section for our candidate $\chi$ are calculated using {\tt micrOMEGAs}. For the latter we assume it depends on the model parameters in a straightforward manner, as we expect for the $\chi\bar{\chi}\rightarrow\mu^+\mu^-$ tree level dominant process,
\begin{equation}
    \left\langle \sigma v \right\rangle _{\chi} = g_\mu^2 g_\chi^2 F(m_\chi,m_{Z'}) \; ,
    \label{sigmavParamID}
\end{equation} 
where $F$ encapsulates the portion of the thermally-averaged cross section that does not depend on the coupling parameters. It is also expected that the relic density is inversely proportional to the annihilation cross section $\left\langle \sigma v \right\rangle _{\rm ann}$. Since $\chi$ can only annihilate to muons or muonic neutrinos, which interact with the $Z'$ boson through the same coupling strength $g_{\mu}$, the annihilation cross section will exhibit the same dependence on the couplings as in Eq.~(\ref{sigmavParamID}). We assume then,
\begin{equation}
    \Omega_{\chi}h^2 = \frac{\Omega^1(m_\chi,m_{Z'}) h^2}{g_\chi^2 g_\mu^2} \; ,
    \label{omegaParamID}
\end{equation}
where $\Omega^1(m_\chi,m_{Z'}) h^2$ is the relic density for given masses and $g_\mu=g_\chi=1$. Substituting equations \eqref{sigmavParamID} and \eqref{omegaParamID} into \eqref{limReescID} and setting $g_\chi = \xi g_\mu$,
    \begin{equation}
        \left(\frac{\Omega^1h^2}{\Omega_{\mathrm{DM}}h^2}\right)^2 \frac{F(m_\chi,m_{Z'})}{g_\mu^4 \xi^2}  \leq \left\langle \sigma v \right\rangle _{\mathrm{lim}} \; .
    \end{equation}
    By calculating the factors $F$ and $\Omega^1h^2$ for given masses and $g_\chi=g_\mu=1$ inside {\tt micrOMEGAs}, using the experimental limits extracted from~\cite{Hoof_2020}, and setting $\Omega_{\mathrm{DM}}h^2=0.120$, we can obtain the exclusion limits in the $m_{Z'}-g_\mu$ plane as in the plots below.

We make a brief comment regarding the constraints coming from cosmic-ray positrons as measured in particular by the PAMELA~\cite{PAMELA:2010kea}  and AMS-02~\cite{AMS:2021nhj} experiments. It was shown in~\cite{Bergstr_m_2013,john2021cosmicray} that DM models with annihilations into $\mu^{+}\mu^{-}$ which reproduce the DM thermal relic density via freeze-out can be strongly constrained by these measurements. In particular, in the recent study of~\cite{john2021cosmicray}, it is stated that DM masses below $\sim$ 150 GeV would be excluded for this channel. There are however large uncertainties stemming from the local dark matter density and local magnetic field strength values used, as well as a full lack of knowledge of the characterization of the galactic magnetic field and its effects in the propagation of charged particles, all of which can lead to softer constrains on the DM masses that could potentially be excluded~\cite{Leane:2018kjk}. Due to these large uncertainties, we will only consider for indirect detection constraints the exclusion limits provided by the $\gamma$-ray observations from the annihilation of DM in dSph galaxies, though there are certainly points of our parameter space that could satisfy the cosmic positron constraints as well.

\subsection{Muon anomalous magnetic moment}
\label{sec:muong2}

Recent measurements at Brookhaven National Laboratory and at Fermilab~\cite{Muong-2:2006rrc, Muong-2:2021ojo} of the muon magnetic moment result in a combined excess of 4.2  $\sigma$ with respect to the SM predicted value (see \cite{Aoyama:2020ynm} and references therein). Due to the axial nature of the $Z'$ coupling to charged muons,  there is a negative one loop contribution to the muon magnetic moment~\cite{DEramo:2017zqw} with respect to the SM contribution, 
\begin{equation}
    \delta a_\mu^{Z'} = -\frac{5}{12\pi^2}\frac{m_\mu^2}{m_{Z'}^2} g_\mu^2\; ,
    \label{NPcontrgmu}
\end{equation}
which worsens the agreement with respect to the measured value. This thus implies a constraint on the possible $Z'$ mass and coupling to the visible sector in our model and indirectly a constraint on the DM as well. We demand that the new physics contribution does not exceed the combined theoretical and experimental uncertainties at the $2\sigma$-level,
\begin{equation}
    \left|\delta a_\mu^{Z'}\right| \leq 2\sqrt{\sigma_{\mathrm{exp}}^2+\sigma_{\mathrm{th}}^2}\; .
\end{equation}
where $\sigma_{{\rm exp}}= 41\times 10^{-11}$ and $\sigma_{{\rm th}}= 43\times 10^{-11} $. Notice that with this requirement the NP contribution of Eq.~(\ref{NPcontrgmu}) is hidden in the theoretical and experimental uncertainties, keeping the disagreement with the measured value below the 5$\sigma$ level.

It is worth mentioning that lattice QCD calculations have been shown to modify the SM theoretical predictions~\cite{Borsanyi_2021}, reducing the discrepancy with respect to the experimental measurements to a  1.6$\sigma$ excess~\cite{Balkin:2021rvh}, which would also reduce the tension within our model. 

\subsection{Trident diagrams}
\label{sec:trident}

The $Z'$ interaction with muonic neutrinos  are constrained by what are known as neutrino trident experiments, which probe lepton production by neutrino scattering with nucleons. In our model, the process $\nu_\mu N \rightarrow \nu_\mu \mu^+ \mu^- N$ takes place with the contribution of 
$Z'$ exchange diagrams. Measurements of the associated cross section at the Columbia-Chicago-Fermilab-Rochester  (CCFR) neutrino experiment at the Fermilab Tevatron~\cite{CCFR:1991lpl, Altmannshofer:2019zhy} imply the following constraint,
\begin{equation}
    2v^2\frac{g_\mu^2}{m_{Z'}^2} \leq 0.6\; ,
\end{equation}
where $v\simeq 246\;\mathrm{GeV}$ is the EW breaking vacuum expectation value. Interestingly enough, for the case of only axial contributions as in our scenario, it is expected that future Deep Underground Neutrino Experiment (DUNE) will have similar sensitivity as CCFR and therefore no improvement of this particular constraint is expected.

\subsection{Collider searches at the LHC}
\label{sec:collider}
\begin{figure}[t]
\centering
\begin{tabular}{ccc}
\includegraphics[scale=0.45]{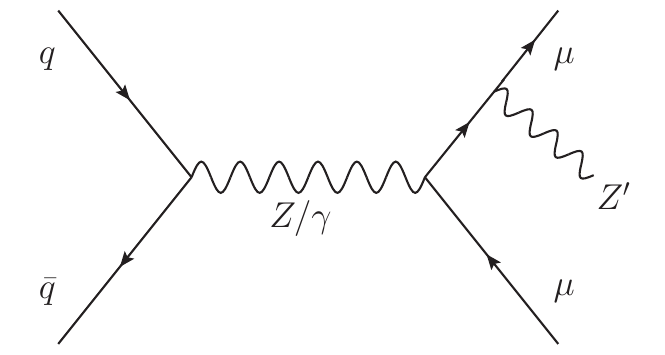}&\includegraphics[scale=0.45]{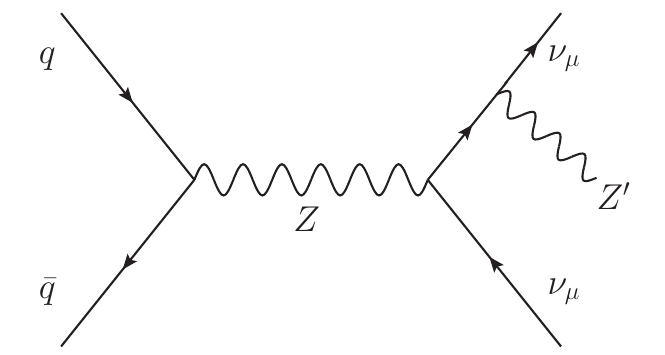} &
\includegraphics[scale=0.45]{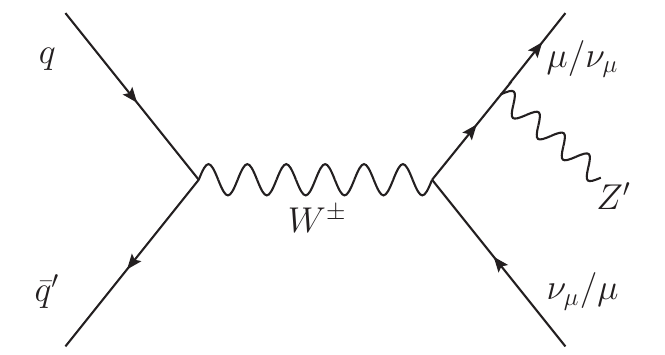}
\end{tabular}
\caption{Leading order diagrams corresponding to the production of the $Z'$ boson at the LHC.}
\label{diagrams}
\end{figure}

At the LHC the leptophilic $Z'$ boson is mainly produced through Drell-Yan processes with the $Z'$ radiated from one of the final leptons, as shown in Fig.~\ref{diagrams}. The specific final state  depends on which is the exchanged EW gauge boson and the decay mode of the $Z'$. For the muonic $Z'$ the exchange of $\gamma$ and $Z$ bosons lead to final states with 4$\mu$ or $2\mu$+ ${E}^{\mathrm{miss}}_T$~\cite{Bell:2014tta}, while the exchange of $W$ gives rise to signals with $3\mu$+${E}^{\mathrm{miss}}_T$ or $1\mu$+${E}^{\mathrm{miss}}_T$. Note that the missing transverse energy in the modes with one or two muons can arise not only from the neutrinos but also from the DM particles. According to the analyses in~\cite{Ma:2001md,delAguila:2014soa,Drees:2018hhs} the constraints coming from the modes with at least three leptons are the most restrictive ones for the case of a leptophilic vector boson. Moreover, it is also shown in~\cite{delAguila:2014soa} that in general the $3\mu$+${E}^{\mathrm{miss}}_T$ mode provides stronger bounds than the 4$\mu$ mode, except when the $Z'$ coupling to right-handed muons ($g_R$) is larger than to left-handed muons ($g_L$), with the limits still being similar for $g_R/g_L$ as large as 2. In our case,  since the $Z'$ is VA coupled to muons, we have $g_R=-g_L$, which translates to cross sections for the $3\mu$+${E}^{\mathrm{miss}}_T$ channel that are $\sim 3$ times those of the $4\mu$ mode along the parameter space explored here.

In order to obtain an estimation of the limits imposed by the existing searches at the LHC, we confront the most promising channel of the model, $3\mu$+${E}^{\mathrm{miss}}_T$, with the collection of LHC analyses implemented in the Public Analysis Database (PAD) of {\tt MadAnalysis5}~\cite{Conte:2014zja,Dumont:2014tja,Conte:2018vmg}. We also include the limits coming from the $2\mu$+${E}^{\mathrm{miss}}_T$ channel since in the model we are considering its cross section increases with $g_{\chi}$ and then it is interesting to test if there is some regime in which it gives better exclusion bounds than the $3\mu$+${E}^{\mathrm{miss}}_T$ mode. The signals of both channels were simulated with {\tt MadGraph 3.1.1}~\cite{Alwall:2014hca} using an implementation of the model in UFO format obtained with {\tt Feynrules}~\cite{Alloul:2013bka}. The parton shower and hadronization were carried out with {\tt Pythia 8.2}~\cite{Sjostrand:2014zea}, while a fast detector simulation is performed inside {\tt MadAnalysis5} by {\tt Delphes 3}~\cite{deFavereau:2013fsa}. In Fig.~\ref{figPAD} we show the exclusion limits in the mass range $(200-500)\;{\rm GeV}$~\footnote{This range reflects the fact that $Z'$ masses lower than the $Z$ boson mass have been already studied in the literature (see, for instance,~\cite{Dror:2017nsg,Ismail:2017fgq}). In addition, we will show in Section~\ref{sec:mucoll} that the analysis of the decay $Z' \rightarrow ZZ$ (with both $Z$ bosons produced on shell) in a muon collider would be crucial in order to probe the anomalous nature of the model. Accordingly, $m_{Z'}$ = 200 GeV is simply taken as a reference lower value. The upper limit of $500\;\mathrm{GeV}$ is chosen \textit{a posteriori} given that the collider searches loose sensitivity.} arising from the most restrictive search of the PAD along with the bounds imposed by the CCFR neutrino experiment and the latest measurements of the muon magnetic moment at Fermilab. We show two representative cases: $g_{\chi}=0$ (left panel) and $g_{\chi}\neq 0 $ with $\mathrm{BR}_{\mathrm{inv}}=0.6$ and $z_{\chi}=0.01$. For the latter we also include limits from indirect detection and relic density. The most restrictive limit for the $2\mu +E^{\mathrm{miss}}_T$ channel corresponds to the inclusive signal region SR\_SF\_1J with $m_{T2}>160$ GeV of the ATLAS search for electroweakinos and sleptons in the $2\ell+{E}^{\mathrm{miss}}_T$ channel~\cite{Araz:2020dlf,ATLAS:2019lff}. For the $3\mu$+${E}^{\mathrm{miss}}_T$ mode the most constraining bound is obtained with the signal region A44 of the CMS search for electroweakinos in multileptons final states~\cite{cms_sus_16_039,CMS:2017moi}. As can be seen from Fig.~\ref{figPAD}, none of these bounds are competitive with those arising from the trident diagrams and the 
muon anomalous magnetic moment. Regarding the comparison between the $2\mu$+${E}^{\mathrm{miss}}_T$ and $3\mu$+${E}^{\mathrm{miss}}_T$ signatures, we see that the latter provides the stronger limits as expected. However, the opening of the DM decay mode of the $Z'$ slightly improves the limit provided by the $2\mu$+${E}^{\mathrm{miss}}_T$ and at the same time worsens the $3\mu$+${E}^{\mathrm{miss}}_T$ bound. This is also expected due to the behavior of the corresponding cross sections when the decay channel $Z'\to \chi\bar{\chi}$ is available. Nevertheless, the gap between the exclusion limits is still significant and we have checked that they become comparable only when $\mathrm{BR}_{\mathrm{inv}}$ is as large as 0.9. For this reason we will focus exclusively on the $3\mu$+${E}^{\mathrm{miss}}_T$ signature in what follows.

The lack of sensitivity of the existing searches at the LHC to exclude and/or discover the model studied here is in part due to the fact that the signal regions mentioned above select events with both electrons and muons in the final state instead of only muons, which would be better suited to probe a muonic $Z'$. Thus, in order to have a better sense of the LHC potential to test the $U(1)'_{\mu}$ axial model it is necessary to consider a dedicated search strategy. Here we apply the one proposed in~\cite{delAguila:2014soa}, which is based on the ATLAS search~\cite{ATLAS:2013rla}. In the following list we summarize the cuts defining the search strategy:

\begin{figure}[t!]
\includegraphics[width=\textwidth]{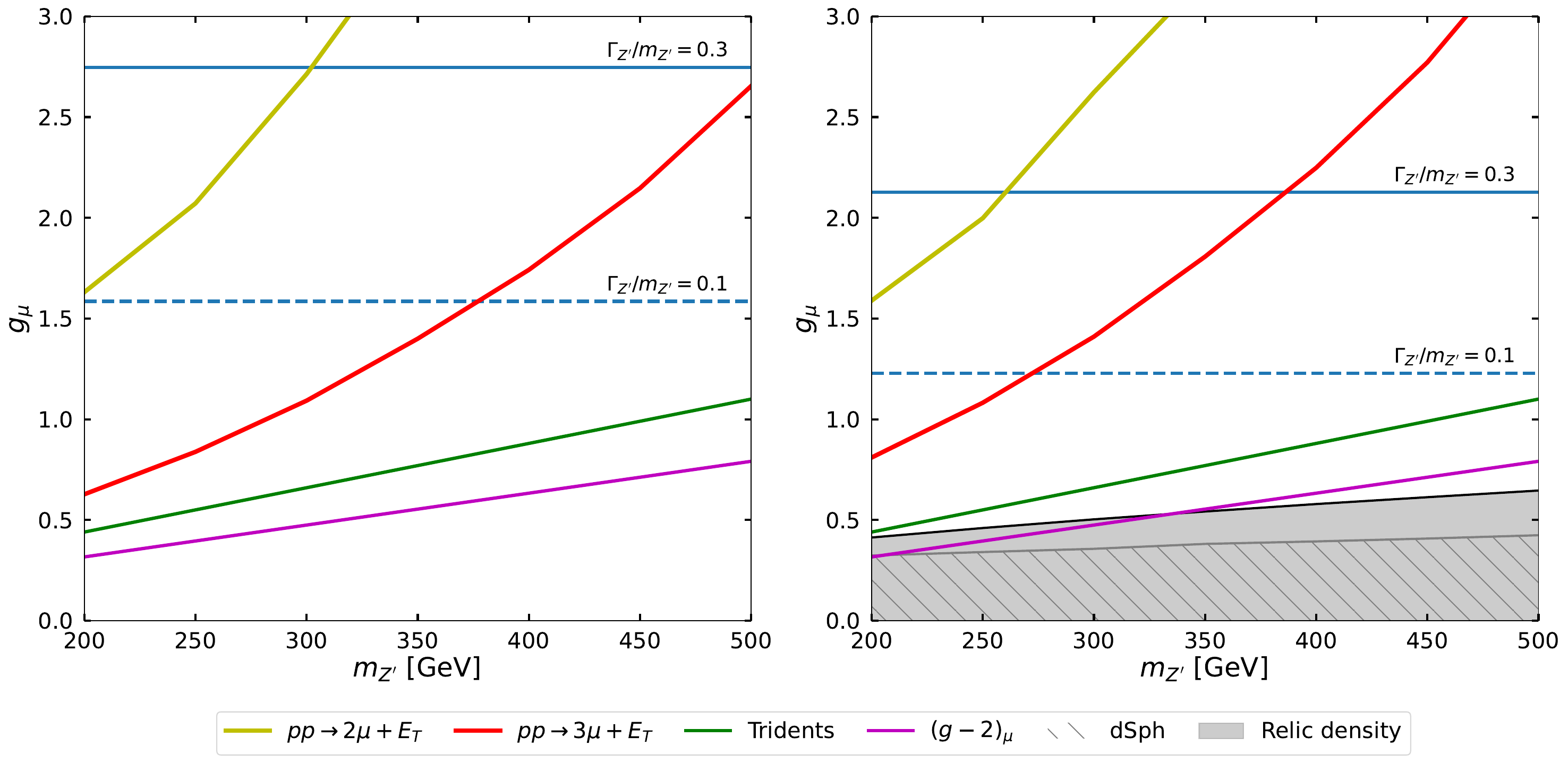}
\caption{Exclusion limits derived from existing LHC searches (see the text for details), the CCFR neutrino experiment and the measurement of the muon anomalous magnetic moment for $g_{\chi}=0$ (left panel) and  $g_{\chi}\neq 0$ with $\mathrm{BR}_{\mathrm{inv}}=0.6$ and $z_{\chi}=0.01$ (right panel). For the latter we also include the regions excluded by the relic density and the dSph galaxies bound from Fermi-LAT. We add for reference the lines corresponding to $\Gamma_{Z'}/m_{Z'}$ ratios 0.1 and 0.3.}
\label{figPAD}
\end{figure}

\begin{enumerate}
\item A set of basic cuts are applied to the muons and jets at generator and detector level: $p^{\mu}_T>50$ GeV, $|\eta_{\mu}|<2.4$, $\Delta R_{j\mu}>0.4$, $p^j_T>20$ GeV and $|\eta_j|<2.5$.
\item A $b$-jet veto is applied ($N_b=0$) and only events with exactly three muons ($N_{\mu}=3$) with net charge $\pm 1$ are kept.
\item Two more vetoes are applied in terms of requirements on the invariant mass of the opposite sign muon pairs: $m_{\mu^+\mu^-}>12$ GeV (low resonance veto) and $|m_{\mu^+\mu^-}-m_Z|>10$ GeV ($Z$ veto).
\item The missing transverse energy is required to be above 100 GeV.   
\item The transverse mass of the muon not belonging to the pair that better reconstructs the $Z'$ mass must be larger than 110 GeV.
\item Finally, at least one opposite sign muon pair must fulfill the requirement $|m_{\mu^+\mu^-}-m_{Z'}|<0.1 m_{Z'}$.
\end{enumerate}

As it is pointed out in~\cite{delAguila:2014soa}, the last cut is crucial to enhance the sensitivity and obtain significant limits. Notice that this cut would not be suitable for the $2\mu+E^{\mathrm{miss}}_T$ channel where part of the signal events arises from the $Z'\mu^+\mu^-$ production followed by the $Z'$ invisible decay, which makes impossible the reconstruction of the $Z'$ boson. This is another reason to prefer the $3\mu+E^{\mathrm{miss}}_T$ channel instead.

In order to obtain the prospects for discovery and exclusion derived from this search strategy, we simulated the dominant backgrounds  using the same simulation setup already described for the simulation of the signal. We considered the diboson processes leading to $\ell\ell\ell'\nu_{\ell'}$ and $\ell\ell\ell'\ell'$ final states, where $\ell,\ell'=\mu,\tau$, and, in addition, the triboson processes $WWW,WWZ,WZZ$ and $ZZZ$ that produce final states with the same charged lepton content. Since the search strategy described above is based on the signal region SRnoZc of~\cite{ATLAS:2013rla}, we used the corresponding background rates to validate our simulation procedure, finding good agreement within the reported uncertainties. The results obtained by applying the dedicated search strategy to the most promising channel will be presented in Section~\ref{Results} along with the experimental constraints discussed previously in this section.

\section{LHC projections for Run 3 and high luminosity}
\label{Results}

In order to estimate the exclusion/discovery prospects for the $U(1)'_{\mu}$ axial model at the LHC with the search strategy described in Section~\ref{sec:collider}, we will consider the four benchmarks listed in Table~\ref{tab1}. Each benchmark is defined by its invisible branching ratio $\mathrm{BR}_{\mathrm{inv}}$ and squared mass ratio $z_\chi=m_\chi^2/m_{Z'}^2$ values, which also fix the coupling ratio $\xi=g_\chi/g_\mu$ in virtue of Eq.~\eqref{eq5}. For each benchmark we generate $pp\rightarrow 3\mu + E_T^{\mathrm{miss}}$ signal events varying the coupling $g_{\mu}$ and the mass $m_{Z'}$, while adjusting $g_\chi$ and $m_\chi$ to the corresponding values that fix $\mathrm{BR_{inv}}$ and $z_\chi$. We use the same UFO model as the one implemented for the PAD recasting described in Section~\ref{sec:collider}. Parton level events are generated using \texttt{MadGraph}, parton shower and hadronization are simulated by \texttt{Pythia 8.2}, and fast detector simulation is done by \texttt{Delphes 3}. Detector-level events are then passed through the analysis described in Section~\ref{sec:collider}, which is implemented  in \texttt{MadAnalysis5}, where we define a single signal region (SR) for the events that pass all cuts. We obtain the detector-level number of events that pass the SR cuts as
\begin{equation}
	n = \mathcal{L}\sigma\mathcal{A}\; ,
	\label{sigmaLumiAcc}
\end{equation}
where $\mathcal{L}$ is the total integrated luminosity, $\sigma$ is the process cross section and $\mathcal{A}$ is the selection acceptance. The latter is calculated as the fraction of events that pass all the cuts over the total number of generated events. Eq.~\eqref{sigmaLumiAcc} is used for both signal and background processes. The 95\% C.L. exclusion limits and the discovery prospects are obtained \cite{Cowan:2010js} by demanding
\begin{equation}
\label{excl}
\mathcal{S}_{\mathrm{excl}} = \sqrt{2\left(b\,\ln\left(\frac{b}{s+b}\right)+s\right)} = 1.64
\end{equation}
and
\begin{equation}
\label{disc}
\mathcal{S}_{\mathrm{disc}} = \sqrt{-2\left((s+b)\,\ln\left(\frac{b}{s+b}\right)+s\right)}= 5,
\end{equation}
respectively, where $s$ and $b$ denote the number of signal and background events that pass the SR cuts. We do not include systematic uncertainties in our analysis.

\renewcommand*{\arraystretch}{1.5}
	\begin{table}[t!]
	\begin{center}
	\begin{tabular}{cccc}
	\hline
	Benchmark & $\mathrm{BR}_{\mathrm{inv}}$ & $z_{\chi}$ & $\xi$ \\
	\hline
	BMI & 0.6 & 0.01 & 1 \\
	BMII & 0.6 & 0.16 & 1.12 \\
	BMIII & 0.9 & 0.01 & 2.92 \\
	BMIV & 0.9 & 0.16 & 3.28 \\
	\hline
	\end{tabular}
	\end{center}
	\caption{Parameters corresponding to the four benchmarks of the $U(1)'_{\mu}$ axial model considered in this section.}
	\label{tab1}
	\end{table}

\begin{figure}[!t]
	\centering
	\includegraphics[width=\textwidth]{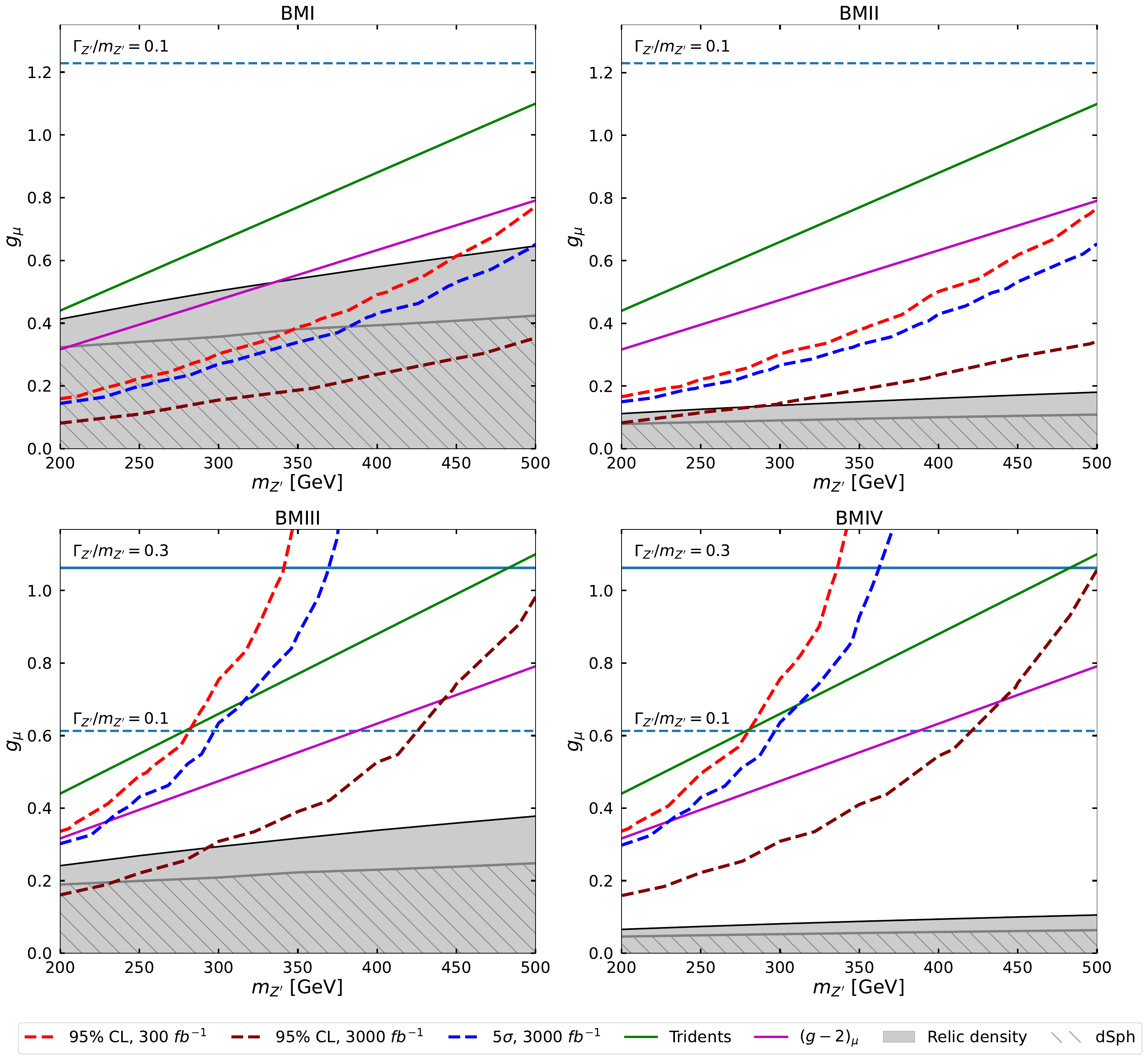}
	\caption{Summary of the constraints evaluated for the $U(1)'_\mu$ model at the four different benchmarks introduced in Table \ref{tab1}. Current available experimental bounds are represented by solid lines. 
	The gray solid regions are excluded by relic density as we mention in Section~\ref{sec:relic}, and gray hashed regions are also excluded by dSph observations as we explain in Section~\ref{sec:indirect}.
	The upper bounds imposed by $(g-2)_\mu$ correction and trident diagrams are described in Secs.~\ref{sec:muong2} and \ref{sec:trident} respectively.
	The LHC limits obtained with the analysis described in Section~\ref{sec:collider} are shown in dashed lines, namely $95\%$ C.L. exclusion limits at luminosities of $300\,\mathrm{fb}^{-1}$ and $3000\,\mathrm{fb}^{-1}$ are represented by red and brown dashed lines respectively, and $5\sigma$ discovery limits at $3000\,\mathrm{fb}^{-1}$ are shown in blue. 
	Horizontal lines represent reference values for $\Gamma_{Z'}/m_{Z'}$.}
	\label{resultados}
\end{figure}

We show in Fig.~\ref{resultados} the exclusion and discovery limits in dashed contours, along with the other experimental constraints described in Section~\ref{sec:expcon}, for the $Z'$ mass range of $(200-500)\;\mathrm{GeV}$. We use $\mathcal{L}=300\;\mathrm{fb}^{-1}$ as the LHC is expected to reach this value in the near future, and $\mathcal{L}=3000\;\mathrm{fb}^{-1}$ to account for the full LHC lifetime. We also plot $\Gamma_{Z'}/m_{Z'}$ reference values, as we restrict our analysis to $\Gamma_{Z'}/m_{Z'}<0.3$.
Note that, for every benchmark considered, there is a portion of the parameter space that remains unconstrained by current experimental data. These available regions are bounded below by the relic density constraint, for which the gray area in the plots represent points that predict a $\chi$ relic density that would overclose the Universe (i.e. $\Omega_{\chi} > \Omega_{\mathrm{DM}}$), and therefore cannot be allowed. On the other hand, upper bounds can be taken up to the more restrictive magenta lines, below which our model predicts a $(g-2)_\mu$ correction that lies within the $2\sigma$ experimental uncertainty, as described in Section~\ref{sec:muong2}. Taking into account that there could be other new physics sources for this quantity that cancel the $Z'$ exchange contribution, but more importantly that there are lattice results~\cite{Borsanyi_2021,Balkin:2021rvh} that show a smaller discrepancy between the SM prediction and experimental values, we could relax the muon anomalous magnetic moment bound in the plots and consider the region below the trident constraint to be explored by our collider analysis.

For the benchmark point BMI there is a region of the parameter space above $m_{Z'}\sim350\;\mathrm{GeV}$ that is unconstrained by current experimental bounds. The $95\%$ C.L. exclusion limits we obtain for the collider analysis at $300\;\mathrm{fb}^{-1}$ would exclude our model up to $450\;\mathrm{GeV}$ for this benchmark. Moreover, the $3000\;\mathrm{fb}^{-1}$ exclusion and discovery prospects cover the available parameter space in the range of masses considered.
The current experimental bounds evaluated at the benchmark point BMII allow for $m_{Z'}$ values along the whole span of masses considered. A portion of this region could be explored at $300\;\mathrm{fb}^{-1}$, improving the $(g-2)_\mu$ limit in the $(200-500)\;\mathrm{GeV}$ range. At $3000\;\mathrm{fb}^{-1}$ for BMII it would be possible to completely exclude our model for $m_{Z'}<300\;\mathrm{GeV}$, and most of the $g_\mu$ range can be excluded for larger masses (for example, with $m_{Z'}=500\;\mathrm{GeV}$, $g_\mu\gtrsim0.35$ is excluded, which corresponds to $\Gamma_{Z'}/m_{Z'}\gtrsim  8\times 10^{-3}$). Discovery prospects at the LHC for this benchmark also cover a portion of the available parameter space in the range of masses considered.
At benchmark point BMIII our $300\;\mathrm{fb}^{-1}$ exclusion limit is weaker than the $(g-2)_\mu$ constraint. As we mentioned before, this constraint can be relaxed, so our collider limit at $300\;\mathrm{fb}^{-1}$ shows an improvement with respect to the trident bound below $280\;\mathrm{GeV}$, becoming no longer competitive for larger masses. The discovery prospect also sits between $(g-2)_\mu$ and trident limits for masses in the $(230-320)\;\mathrm{GeV}$ range, while being slightly below $(g-2)_\mu$ bound for lower masses. The $3000\;\mathrm{fb}^{-1}$ exclusion limit we obtain fully covers the available parameter space for $m_{Z'}<300\;\mathrm{GeV}$, and improves the $(g-2)_\mu$ limit for masses up to $430\;\mathrm{GeV}$, where it excludes $g_\mu\gtrsim0.7$ ($\Gamma_{Z'}/m_{Z'}\gtrsim 0.13$).
Collider limits in benchmark point BMIV show a similar behavior to BMIII. For example, at $m_{Z'}=200\;\mathrm{GeV}$, $g_\mu\gtrsim0.17$ ($\Gamma_{Z'}/m_{Z'}\gtrsim 7.7\times10^{-3}$) is excluded at the HL-LHC, while for $m_{Z'}=430\;\mathrm{GeV}$ it excludes $g_\mu\gtrsim0.7$ as before. 
However, since lower masses are allowed by the relic density bound in comparison with BMIII, there is a larger unconstrained region in this benchmark in the whole $(200-500)\;\mathrm{GeV}$ mass range, even at $3000\;\mathrm{fb}^{-1}$.

Note that for benchmark points BMI and BMIII the relic density lower bound is higher than in BMII and BMIV. This is due to the fact that their $z_\chi$ value is lower, further away from the resonant production, thus needing a larger coupling to reproduce the annihilation cross section value associated with the measured DM relic density. The same reasoning applies to the dSph constraints, as they are also related to the $\chi\bar{\chi}$ annihilation to SM states via $Z'$ (see Section~\ref{sec:indirect}). 
Also, collider limits obtained for BMI and BMII are stronger since the $Z'$ branching ratio to muons, i.e. $1-\mathrm{BR_{inv}}$, is larger. In these scenarios it would be possible to improve the bounds for our model in the near future at the LHC. For a $Z'$ with larger $\mathrm{BR_{inv}}$ (BMIII and BMIV) one would need the estimated luminosity of the full LHC lifetime to probe a significant portion of the available parameter space, while still leaving some areas inaccessible to the LHC. 
Finally, it is worth noting that collider limits shown for BMIII and BMIV become steeper for higher masses and couplings. This is due to the fact that $\Gamma_{Z'}$ becomes larger for these values, as we can see in Eq.~\eqref{eq6}. Therefore, the mass window cut (cut 6 of the analysis in Section~\ref{sec:collider}) becomes less effective as the width of the resonance increases and more signal events are left out of the signal region. We tried to substitute the mass window cut with other similar cuts that attempt to include more signal muons, but we only observed a small improvement in the limits for a ``wider window", using the condition $|m_{\mu^+\mu^-}-m_{Z'}|<0.3\, m_{Z'}$ instead of cut 6. Nonetheless, the regions where the wider window limits are competitive with the ones in Fig.~\ref{resultados} are above trident and/or $(g-2)_\mu$ bounds, and thus are already excluded. It is also worth noting that in all of the cases described the limits in which we include a $m_{\mu^+\mu^-}$ window around the $m_{Z'}$ value are stronger than the ones obtained without a mass window. We show some examples of cutflows for this analysis in the Appendix~\ref{appendix:b}.

\section{Mixed gauge anomalies}\label{sec:anomalies}

In the model under consideration the axial (chiral) coupling of the muon (muonic neutrino) to the $Z'$ boson induce mixed anomalies between $U(1)'_{\mu}$ and the EW gauge symmetry group and gravity. Gauge symmetries of the classical fields cannot be simultaneously satisfied in the QFT and this is signaled by  anomalous Ward identities which imply the loss of unitarity and/or Lorentz invariance in the first place and the non-renormalizabilty of the theory in second place~\cite{PRESKILL1991323}. By allowing the gauge field to acquire mass (as we are doing via spontaneous symmetry breaking) one naturally solves the problems of unitarity and Lorentz invariance but the issue of renormalizability remains and implies that the theory can only be regarded as an effective theory with a cutoff that cannot be made arbitrarily large without suffering a loss of calculability~\cite{PRESKILL1991323}. In a fully consistent QFT all anomalies must be canceled and in fact, anomalous theories can be regarded as effective models where the anomaly cancellation fully happens at a higher energy scale and in which part of the spectrum responsible for anomaly cancellation has been integrated out in the effective theory~\footnote{For the case of an abelian gauge group with charges covering a vast range of values the validity of the effective model can be pushed up to higher energies and the cutoff depends on the anomaly~\cite{PRESKILL1991323}.}. 

In the case of our anomalous $U(1)'_{\mu}$ symmetry, the Ward identities involving the $U(1)'_{\mu}$ and EW gauge symmetry groups in the EW unbroken phase can be accommodated such that,
\begin{align}
p_{1\mu}\Gamma_{A^{a}A^{b}Z'}^{\mu\nu\rho} & =0,  \label{p1} \\
p_{2\nu}\Gamma_{A^{a}A^{b}Z'}^{\mu\nu\rho} & =0,\label{p2}  \\
(p_{1}+p_{2})_{\rho}\Gamma_{A^{a}A^{b}Z'}^{\mu\nu\rho} & =\frac{\mathcal{A}_{A^{a}A^{b}Z'}}{4\pi^{2}}\epsilon^{\lambda\mu\nu\sigma}p_{1\lambda}p_{2\sigma}, \label{WardZp}
\end{align}
where the $SU(2)_L \times U(1)_Y$ gauge bosons are labeled collectively as $A^i$ and $p_1$, $p_2$ are their momenta. The vertex functions $\Gamma^{\mu\nu\rho}_{A^{a}A^{b}Z'}$ corresponding to triangle diagrams with $Z'$, $A^{a}$ and $A^{b}$ in the external legs are not invariant under a shift of the momenta of the fermions running in the loop. A particular shift to restore the Ward identities for the SM gauge group can be always chosen leading to Eqs.~\eqref{p1}, \eqref{p2}. However, there is no possible shift that satisfies simultaneously the three Ward identities and thus Eq.~\eqref{WardZp} remains non-vanishing. The coefficients of the mixed anomalies between the $Z'$ and the SM gauge bosons can be calculated as,
\begin{equation}
    \mathcal{A}_{A^{a}A^{b}Z'} = \frac{1}{2} \left[
    \mathrm{Tr}_{\mathcal{R}_R} (Q'\{ T^a,T^b \}) - \mathrm{Tr}_{\mathcal{R}_L} (Q'\{ T^a,T^b \} )
    \right]\; ,
\label{Coeff}
\end{equation}
with $Q'$, $T^i$ being the $U(1)'_{\mu}$ and EW gauge group generators, respectively, and $\mathrm{Tr}_{\mathcal{R}_{R(L)}}$ standing for the symmetric part of the trace evaluated in the right-handed (left-handed) chiral representation of the SM fermions running in the loop. The anomalous coefficients $\mathcal{A}_{Z'Z'B}$, $\mathcal{A}_{Z'Z'W^a}$ and $\mathcal{A}_{Z'BW^a}$ are zero because the EW generators are traceless and factor out in Eq.~\eqref{Coeff}. The non-vanishing anomalous coefficients are,
\begin{equation}
\mathcal{A}_{Z'BB} = \frac{3}{2}Q_\mu \, , \; \; \; \; \mathcal{A}_{Z'W^aW^b} = \frac{\delta_{ab}}{2}Q_\mu \, , \; \; \; \; \mathcal{A}_{Z'Z'Z'} = 3Q_\mu^3 \, .\label{anom1}
\end{equation}
The presence of 1-loop anomalous triple gauge couplings~\cite{Dedes:2012me}, in particular the ones involving the $Z'$ and the EW gauge bosons, renders the proposed model interesting from a phenomenological perspective. Indeed, we can consider these interactions as remnants at low energy of the UV physics which completes our effective theory and thus, if we were able to probe these couplings, we would be having access through a window into this UV physics. 
We shall discuss in Sections~\ref{sec:anomcolliderhadron} and~\ref{sec:mucoll} the potential reach of searches for anomalous couplings at the LHC and at hypothetical 100 TeV proton and muon colliders.

\subsection{Example of an anomaly-free UV completion}

We briefly discuss now a possible anomaly-free UV completion of our effective model. With this aim we introduce new fermions in order to cancel all the gauge and gravity anomalies in the high energy theory. Following~\cite{Batra:2005rh} it is possible to construct an anomaly-free extension by introducing a minimum of four 2-dimensional Weyl fermions that transform under $SU(3)_c\times SU(2)_L$ and have charges under $U(1)_Y \times U(1)'_{\mu}$ as,
\begin{equation}
\begin{aligned}
    \psi_{L}^{l}&:\left(\mathbf{1},\mathbf{2},-1/2,Q_\mu\right)\,,\\
	\psi_{R}^{l}&:\left(\mathbf{1},\mathbf{2},-1/2,0\right)\,,\\
	\psi_{L}^{e}&:\left(\mathbf{1},\mathbf{1},-1,0\right)\,,\\
	\psi_{R}^{e}&:\left(\mathbf{1},\mathbf{1},-1,-Q_\mu\right)\,.
\end{aligned}
\end{equation}
After the $U(1)'_{\mu}$ symmetry is spontaneously broken at a scale above the EW scale, the left-handed and right-handed Weyl fermions that transform in the same way under the SM gauge group combine to form a 4-dimensional Dirac fermion
 with only vectorial couplings to the SM gauge bosons, therefore avoiding the introduction of anomalies associated with the SM gauge group. Furthermore, their charges under $U(1)'_{\mu}$ and their chiral nature allows them to have  masses generated by the spontaneous breaking of the $U(1)'_{\mu}$ symmetry through the vev of the $\Phi$ Higgs field (see Section~\ref{sec:model}). Under the assumption of Yukawa couplings of order one, the spectrum of the new fermions naturally lies at the $U(1)'_{\mu}$ breaking scale $f$. Notice that since the new fermions are not charged under the color group they are not strongly constrained by direct searches at the LHC and could have masses of order $\sim$ few TeV, which at the same time implies that the $U(1)'_{\mu}$ breaking scale $f$ needs not to be pushed up to values far above the TeV scale.

\subsection{Hadron collider searches for anomalous triple gauge boson couplings}\label{sec:anomcolliderhadron}

As it has been discussed throughout the text, our model represents an anomalous EFT which can be interpreted as a UV complete theory in which part of the chiral fermion spectrum charged under the $U(1)'_{\mu}$ has been integrated out. This implies in particular an enhancement with the energy of the process in the coupling between the longitudinal mode of the $Z'$ and two EW gauge bosons~\cite{Dror:2017nsg,Ismail:2017fgq}, as can be seen from the anomalous Ward identities, signaling the breaking of unitarity. Such behavior of course is tamed at high energies by the appearance of the spectator fermions and the restoration of a full unitary theory in the UV, though at intermediate energies the enhancement remains and could potentially soften the 1-loop suppression of the anomalous couplings, providing hope of probing these at current and future particle accelerators. 

\begin{figure}[t]
\centering
	\begin{subfigure}[t]{0.45\textwidth}
		\centering
		\includegraphics[]{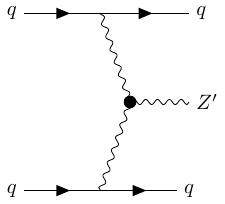}	
		\caption{Anomalous Vector-boson fusion \label{fig:vbf}}	
	\end{subfigure}	
	\hfill
	\begin{subfigure}[t]{0.45\textwidth}
		\centering
		\includegraphics[]{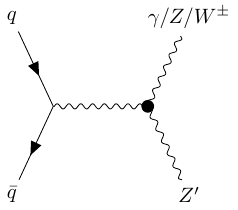}		
		\caption{Anomalous production in the s-channel. \label{fig:anomcanals}}
	\end{subfigure}	
\caption{Anomalous production diagrams at hadron colliders. The internal vector boson lines correspond to $\gamma$, $Z$ o $W^\pm$ accordingly and the full dot represents the anomalous triple gauge coupling. \label{diagscol}}
\end{figure}

In Fig.~\ref{diagscol} we show two situations in which the anomalous triple gauge coupling could enter in the production of a $Z'$ at hadron colliders. The first one corresponds to a vector boson fusion process and as such, due to the t-channel production of the $Z'$, we expect no enhancement with the energy in the process. The second process on the other hand is an s-channel $Z'$-strahlung production, where the $Z'$ is emitted from an EW gauge boson in the s-channel. In this case, due to the s-channel production of the intermediate EW gauge boson,  the $Z'$ can potentially be sufficiently boosted such that the enhancement with energy in the production of the longitudinal mode of the $Z'$ leads to testable cross sections. Since we expect the longitudinal mode of the $Z'$ to dominate the production cross sections in the high energy regime and in order to avoid the complications from having to integrate over the anomalous momentum-dependent couplings convoluted with the proton's PDF, we can further simplify our calculations using the Goldstone equivalence theorem to estimate the amplitude of the process of interest,
\begin{equation}
    \mathcal{M}(Z'_{L},\dots)=\mathcal{M}(\phi',\dots)+\mathcal{O}\left(\frac{m_{Z'}}{\sqrt{s}}\right) \; .
\end{equation}
where $L$ here denotes the longitudinal polarization of the $Z'$ and $\phi'$ is the corresponding eaten Goldstone boson. We can obtain the anomalous coupling of the Goldstone boson to the EW gauge bosons by replacing in the tree-level Lagrangian Eq.~\eqref{eq1} the interaction term between the $Z'$ and the fermionic currents,
\begin{equation}
    g_\mu Z'_\mu J^\mu_{Z'} \rightarrow \frac{1}{f}(\partial_\mu \phi') J^\mu_{Z'}\; ,\label{eqcurr}
\end{equation}
where $f$ is the Goldstone decay constant which coincides with the vev of $\Phi$. Integrating by parts, we can use that the divergence of the anomalous fermion current is related to the anomaly as~\cite{Bilal:2008qx},
\begin{equation}
\partial_{\mu}J^{\mu}_{Z'}=\frac{  \mathcal{A}_{Z' A^{b}A^{c}}}{32\pi^2}\epsilon^{\mu\nu\rho\sigma}F^{b}_{\mu\nu}F^{c}_{\rho\sigma}\; ,
\end{equation}
where in our case the field strengths correspond to the EW sector $SU(2)_L\times U(1)_Y$. Replacing in Eq.~\eqref{eqcurr} and using the values of Eq.~\eqref{anom1} we obtain,
\begin{equation}
\phi'\left(\frac{3}{64\pi^2}\frac{g_1^2Q_\mu}{f}\epsilon^{\mu\nu\rho\sigma}B_{\mu\nu}B_{\rho\sigma}+\frac{1}{64\pi^2}\frac{g_2^2Q_\mu}{f}\epsilon^{\mu\nu\rho\sigma}W^a_{\mu\nu}W^a_{\rho\sigma}\right)\; ,
\end{equation}
where $g_1$ and $g_2$ are the $U(1)_Y$ and $SU(2)_L$ gauge couplings. Rewriting the last expression in the EW broken physical base  ($\gamma, W^\pm, Z$) we get,  
\begin{equation}
\begin{aligned}
  \frac{\phi'Q_\mu}{32\pi^2 f}  (& 3g_1^2B_{\mu\nu}\tilde{B}^{\mu\nu}+g_2^2 W^a_{\mu\nu}\tilde{W}^{a\mu\nu}) = \\  \frac{\phi'Q_\mu}{32\pi^2 f} \big[ & 4e^2 F_{\mu\nu}\tilde{F}^{\mu\nu} + (3\,\mathrm{tan}^2\theta_W +\mathrm{cot}^2\theta_W)\, e^2 Z_{\mu\nu}\tilde{Z}^{\mu\nu} + 2 (\mathrm{cot}\,\theta_W - 3 \, \mathrm{tan} \, \theta_W )\, e^2 F_{\mu\nu} \tilde{Z}^{\mu\nu} + \\ & + 2 g_2^2 W^+_{\mu\nu} \tilde{W}^{- \,\mu\nu} + 4 i \, s_W g_2^3 (\tilde{W}^{+ \, \mu\nu}A_\mu W^-_\nu + \tilde{W}^{- \, \mu\nu}W^+_\mu A_\nu + \tilde{F}^{\mu\nu} W^-_\mu W^+_\nu ) \, + \\ & + 4 i \, c_W g_2^3 (\tilde{W}^{+ \, \mu\nu}Z_\mu W^-_\nu + \tilde{W}^{- \, \mu\nu}W^+_\mu Z_\nu + \tilde{Z}^{\mu\nu} W^-_\mu W^+_\nu ) \big]   \; ,
  \label{eq:anomPhys}
\end{aligned}
\end{equation} 
where $s_W$ y $c_W$ are the sine and cosine of the Weinberg angle and for each vectorial field $C_{\mu}$ we define  $C_{\mu\nu}=\partial_\mu C_\nu - \partial_\nu C_\mu$ and its dual $\tilde{C}^{\mu\nu} = \epsilon^{\mu\nu\rho\sigma}C_{\rho\sigma}/2$. Note that  effective vertices between $\phi'$ and two or three EW gauge bosons are generated, while vertices between $\phi'$ and four gauge bosons vanish as a consequence of pentagonal diagrams being not anomalous~\cite{Bilal:2008qx}.

The three and four boson anomalous couplings given in Eq.~\eqref{eq:anomPhys} are then implemented in a UFO model via {\tt Feynrules}, which we use to calculate parton-level cross sections using {\tt MadGraph}. We simulate the $pp\rightarrow \phi' V$ signals~\footnote{Four-boson anomalous vertices lead to the possibility of $pp\rightarrow \phi' V V'$ processes, but since these terms are suppressed by a factor of $g_2$ we focus on the triboson vertices only.} as in diagram~\ref{fig:anomcanals}, with $V=\gamma,Z,W^{\pm}$ and the model parameters fixed at $g_\mu = 0.46$, $g_\chi=0.1$, $m_{Z'}=200\;{\rm GeV}$, $m_\chi=30\;{\rm GeV}$ (that correspond to $\xi\approx 0.22$, $z_{\chi}\approx 0.06$, $\mathrm{BR_{inv}\approx 0.35}$), which satisfy $\Omega_\chi h^2 \approx 0.1$ and sit at the trident bound described in Section~\ref{sec:trident}, aiming to maximize the anomalous couplings.
We get the largest cross section for the $pp\rightarrow \phi' \gamma$ channel, resulting in $\sigma_{\phi'\gamma} = 5.9\times 10^{-3}\;{\mathrm{fb}}$, leading to an approximate of $20$ events produced at the LHC at $\sqrt{s}=14\;\mathrm{TeV}$ and at $\mathcal{L}=3000\;\mathrm{fb}^{-1}$, which is estimated for its full lifetime. The cross sections found for the $Z$ and $W$ channels are smaller by a factor of at least $3$, resulting in less than $10$ events. For a dedicated search of this signal it is necessary to impose cuts on the final state particles, further reducing the number of expected events, so probing these anomalous vertices is extremely challenging at the LHC.
In a hypothetical $\sqrt{s}=100\;\mathrm{TeV}$ hadron collider we obtain $\sigma=7.6\times 10^{-2}\;{\mathrm{fb}}$, giving $\sim 230$ events by assuming an integrated luminosity of $3000\;\mathrm{fb}^{-1}$. This might be a more promising scenario for probing the anomalous vertices, though the backgrounds are also expected to grow, but a much more favorable experimental facility for our model would be a muon collider, in which we focus the next section.

\subsection{Muon collider resonant searches for anomalous triple gauge boson couplings}
\label{sec:mucoll}

Muon colliders provide a great potential to explore new physics in the sub-TeV  to the multi-TeV energy range~\cite{AlAli:2021let,Huang:2021nkl,Yin:2020afe}. The large mass of the muon in comparison with the electron mass suppresses synchrotron radiation by roughly a factor of $10^{9}$ for beams of the same energy, and therefore rings can be used to accelerate muon beams efficiently and bring them repeatedly into collision. Furthermore, the physics reach of a muon collider extends that of a proton-proton collider of the same energy since all of the beam energy is available for the hard collision, whereas a fraction of the proton-beam energy is carried by the colliding partons~\footnote{In fact a 14 TeV muon collider provides an effective energy similar to that of a 100 TeV proton-proton collider.}. A dedicated muon collider can scan the Higgs resonance and precisely measure its mass and width~\cite{Barger:1995hr, Barger:1996jm, Han:2021lnp}. In fact, a muon collider is ideal to search for new physics and for resolving narrow resonances both as a precision and/or as an exploratory machine. There are nonetheless challenges that arise from the short muon lifetime and the difficulty of producing large numbers of muons, which requires the development of demanding technologies and new concepts. The beam background from the muon's decay has also consequences on both machine and detector design. 

An ambitious research and development program is needed to assess the feasibility of a muon collider in the tens of TeV range~\cite{Delahaye:2019omf, Bartosik:2020xwr}. Therefore it will be important to study the physics potential of smaller-scale machines in the sub-TeV range that may be built along the way as technology demonstrators. In this respect, a sub-TeV muon collider becomes the ideal machine not only to measure with starking precision by resonant production the main decays of the $U(1)'_{\mu}$ $Z'$ into muon pairs or into invisible particles (muonic neutrinos or DM), but moreover, it may even be possible to probe the triple gauge couplings stemming from the mixed anomalies. 

Before exploring the possibility of probing the triple gauge couplings at a muon collider, we first provide an estimation at Monte Carlo truth level of the discovery prospects for our model given by processes at tree level. In particular, we study two channels which would give the most promising signals: the dimuon channel $\mu^+\mu^-\rightarrow\mu^+\mu^-$ and the monophoton channel $\mu^+\mu^-\rightarrow\gamma+E_T^{{\rm miss}}$. 
For each of these processes we generate signal and background events at parton level with \texttt{MadGraph}. 
As mentioned before, it is expected that a sub-TeV muon collider starts operating at center of mass energy values near the SM Higgs mass, so we fix $\sqrt{s}=125\;{\rm GeV}$ in our simulations. 
This implies that, for the range of $Z'$ masses considered, the channels studied are mediated by an off-shell $Z'$.
We estimate the luminosity required to achieve a $5\sigma$ signal significance for discovery (see Eq.~\eqref{disc}).

\begin{figure}[t]
	\centering
		\begin{subfigure}[t]{0.48\textwidth}
		\centering
			\includegraphics[width=\textwidth]{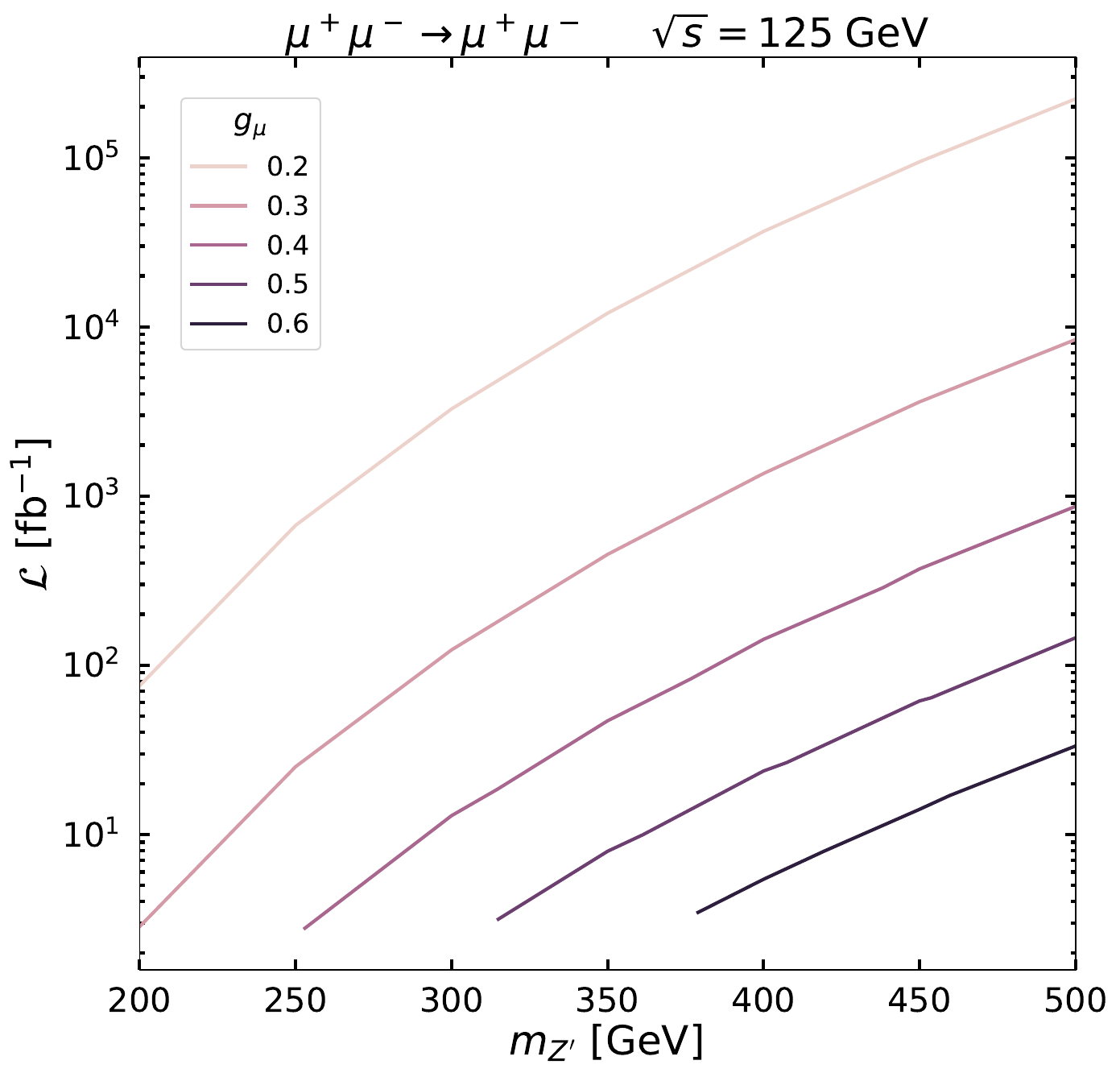}
			\caption{Dimuon channel\label{fig:muon_tree_dimuon}}	
		\end{subfigure}	
		\hfill
		\begin{subfigure}[t]{0.48\textwidth}
			\includegraphics[width=\textwidth]{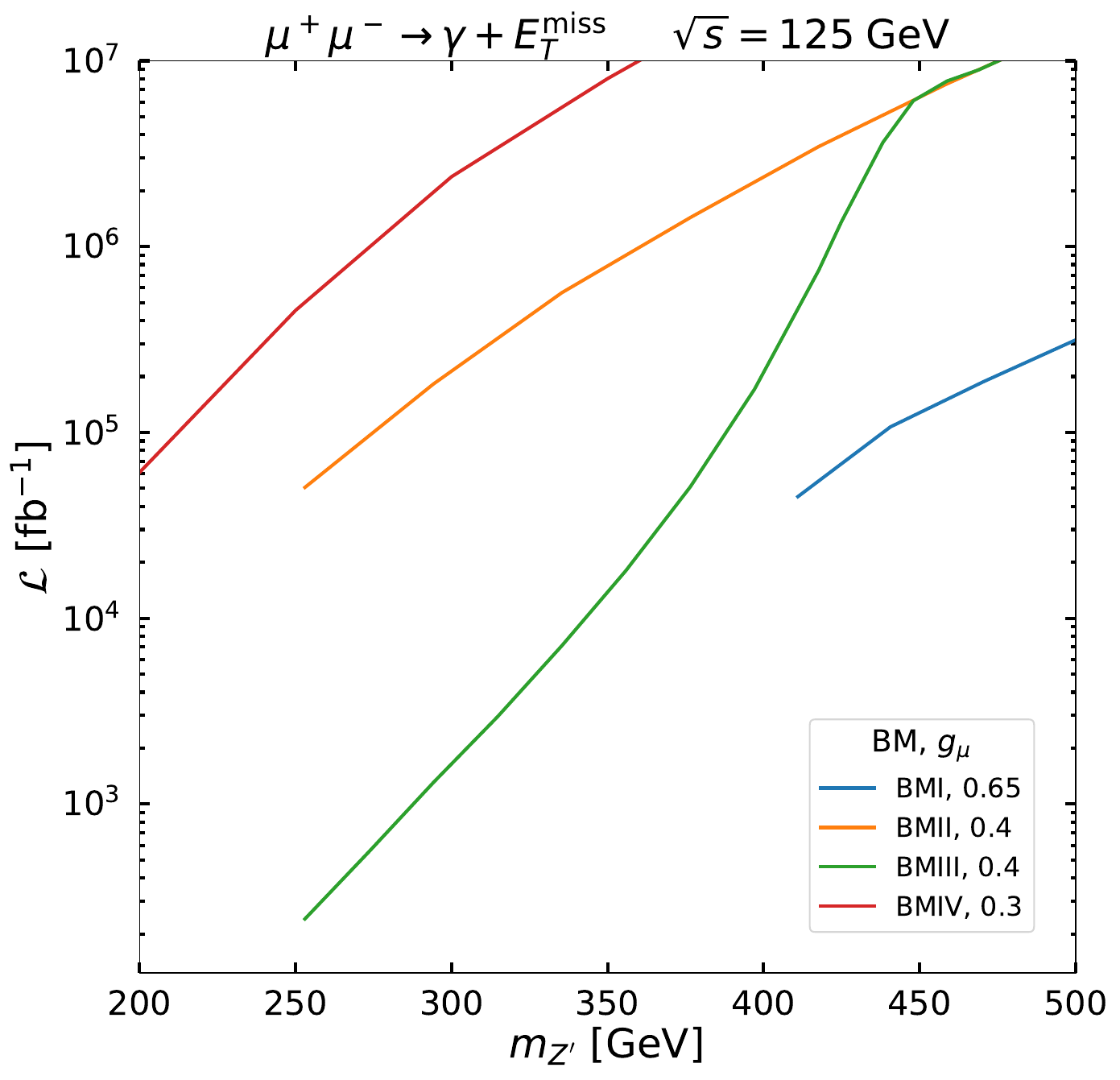}
			\caption{Monophoton channel\label{fig:muon_tree_monophoton}}
		\end{subfigure}	
	\caption{Luminosity required for the discovery of tree level processes at a muon collider with $\sqrt{s}=125\;{\rm GeV}$, for selected points in the parameter space of our model.}
	\end{figure}

In Fig.~\ref{fig:muon_tree_dimuon} we show the required luminosity for the discovery of the dimuon production process at different $g_\mu$ and $m_{Z'}$ values\footnote{We find that the signal cross section is approximately independent of the $g_\chi$ and $m_\chi$ values, due to the off-shell nature of $Z'$ in this diagram.}. 
We impose the {\tt MadGraph} default transverse momentum cut for leptons, $p_T^\ell>10\;{\rm GeV}$, which is consistent with pre-selection cuts applied on full simulations of muon colliders \cite{Zaza:2021bj}.
On the plot we only show points with $(m_{Z'}, g_\mu)$ values that are unconstrained by $(g-2)_\mu$ measurements.
For integrated luminosities of the order of a few tens of ${\rm fb}^{-1}$, discovery would be possible for $g_\mu = 0.6$ in the available region of the $m_{Z'}$ range considered. Furthermore, at $\mathcal{L}\sim 300\;{\rm fb}^{-1}$, discovery can be achieved for: $g_\mu=0.5$ in the available $m_{Z'}$ range, $g_\mu=0.4$ with $m_{Z'}\lesssim 440\;{\rm GeV}$, or even $g_\mu=0.3$ and $m_{Z'}\lesssim 325\;{\rm GeV}$. Note that for $\mathcal{L}\sim\mathcal{O}({\rm few})\;{\rm ab}^{-1}$ one could even probe couplings as small as $g_\mu=0.2$ up to masses of order $m_{Z'}\lesssim 250\;{\rm GeV}$.

For the case of the monophoton channel, we show in Fig.~\ref{fig:muon_tree_monophoton} the target luminosities for the benchmark points provided in Table~\ref{tab1}. For each of these benchmarks we fix $g_\mu$ in order to maximize the $m_{Z'}$ range that is unconstrained by the relic density and $(g-2)_\mu$ bounds from Fig.~\ref{resultados}. Since the dominant background of this channel is the on-shell $\gamma Z$ production with invisible $Z$ decay, and therefore the photon energy is fixed at $E_\gamma\approx 29\;{\rm GeV}$, we impose a cut in the transverse photon momentum $p_T^\gamma>30\;{\rm GeV}$ in order to suppress this background contribution. 
We find that the only realistic scenario for discovery of this channel is BMIII, where the $z_\chi$ value is low enough to keep the $\chi\bar{\chi}$ production open at $m_{Z'}\lesssim 450\;{\rm GeV}$ ($m_\chi\lesssim 45\;{\rm GeV}$), and at the same time the relic density and $(g-2)_\mu$ bounds allow for a wide $m_{Z'}$ range to be probed. Even in this case, only $Z'$ masses below $350\;{\rm GeV}$ can be discovered with $\mathcal{L}<10\;{\rm ab}^{-1}$.
Note that the BMI curve would show a similar behavior as BMIII, with a considerable decrease in the target luminosity at lower masses, but since $(g-2)_\mu$ 
excludes $m_{Z'}\lesssim 400\;{\rm GeV}$ at $g_\mu=0.65$ these points are not shown. 
On the other hand, BMII and BMIV have $m_\chi>80\;{\rm GeV}$, which does not allow DM in the final state at $\sqrt{s}=125\;{\rm GeV}$, and therefore the signal cross section is significantly reduced, depending only on $m_{Z'}$ and $g_\mu$. In fact, this effect can be seen by comparing the BMII and BMIII curves, both at $g_\mu=0.4$: when the DM channel opens for masses $m_{Z'}\lesssim 450\;{\rm GeV}$ in BMIII the required luminosity rapidly drops for lower masses in comparison to BMII, in which only neutrinos contribute to the invisible decay channel. The luminosities required for discovery at benchmarks I, II and IV are above $60\;{\rm ab}^{-1}$, which is considered to be too large for a muon collider, in particular for one with the fixed energy value of $\sqrt{s}=125\;{\rm GeV}$.

Comparing both channels it is clear that the sensitivity for the monophoton channel is significantly lower than the dimuon channel.
If a hypotetical sub-TeV muon collider is able to perform a collision energy scan, it could eventually probe energies that allow on-shell $Z'$ production, which would significantly improve the discovery prospects for both channels in the mass range considered above. A muon collider is also ideal to probe $m_{Z'}$ values higher than $500\;{\rm GeV}$, where the hadronic collider searches at the $14\;{\rm TeV}$ LHC lose sensitivity. 

In the following, we work in a region of parameter space under the assumption that the $Z'$ gauge boson has been already discovered at any of the possible experiments in which it can be searched for and focus on probing the triple gauge couplings stemming from the mixed anomalies, showing that indeed this could be achievable by resonantly producing the $Z'$ at a future muon collider~\footnote{Note that, from a practical point of view, an already discovered $Z'$ makes the aimed center of mass energy for the muon collider known and therefore its construction easier, avoiding having to scan over the energy of the colliding muons.}.

\begin{figure}[t]
	\centering		
	\includegraphics[]{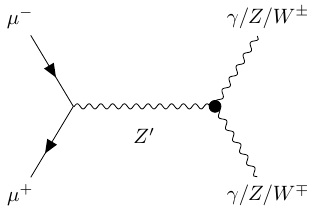}
	\caption{$s$-channel diagram for the production of a $Z'$ boson at muon colliders. The black blob represents the triple gauge couplings arising from the mixed anomalies.}
	\label{fig:musch}
	\end{figure}

Close to the resonance, the cross section corresponding to the $s$-channel production of a $Z'$ boson decaying into a two-body final state consisting of particles $C$ and $D$, can be written as
\begin{equation}
\label{xsmuon}
\sigma_{Z'}(\sqrt{\hat{s}})=\frac{12\pi\,m^4_{Z'}}{(m^2_{Z'}-(m_C+m_D)^2)(m^2_{Z'}-(m_C-m_D)^2)}\frac{\Gamma(Z'\to\mu^+\mu^-)\,\Gamma(Z'\to CD)}{(\hat{s}-m^2_{Z'})^2+m^2_{Z'}\Gamma^2_{Z'}},
\end{equation}
where $\hat{s}=(p_{\mu^+}+p_{\mu^-})^2$ is the center-of-mass energy (c.m.) squared for a given $\mu^+\mu^-$ annihilation and $\Gamma_{Z'}$ is the total decay width of the $Z'$. Here we are interested in analyzing the sensitivity to the anomalous decay channels so that in principle $CD$ can be $\gamma\gamma,Z\gamma,ZZ$ or $W^+W^-$\footnote{Note that in contrast with the anomalous production at hadron colliders, the resonant production of the $Z'$ at a muon collider and its subsequent anomalous decay lead to the simplification of not having to integrate the momentum dependent couplings convoluted with the proton's PDF.}, see Fig~\ref{fig:musch}. However, since we are interested in the resonant production of the $Z'$,  by virtue of the Landau-Yang theorem, the width of the $\gamma\gamma$ channel is zero, and so the resonant cross section  vanishes for this channel. On the other hand, we
expect the prospects of the $W^+W^-$ channel to be similar to those of the $ZZ$ channel, so that we  focus in what follows on the $Z\gamma$ and $ZZ$ final states.

By approximating the energy spectrum of each beam by Gaussian shapes, the effective cross section at the muon collider, $\bar{\sigma}_{Z'}(\sqrt{s})$, can be computed by convoluting $\sigma_{Z'}(\sqrt{\hat{s}})$ with the Gaussian distribution in $\sqrt{\hat{s}}$ centered at $\sqrt{s}$ and with a standard deviation given by~\cite{Barger:1996jm},
\begin{equation}
\label{stddev}
\sigma_{\sqrt{s}}=R\sqrt{s}/\sqrt{2},
\end{equation}
where $R$ is the resolution in the energy of the muon beams. For a proton driver muon facility this resolution is around $0.004$\% for c.m. energy of 126 GeV and increases up to $0.1$\% in the multi-TeV range~\cite{Delahaye:2019omf}. Since we explore $Z'$ masses between 200 and 500 GeV, we will take the former as reference value. The ratio $\Gamma_{Z'}/m_{Z'}$ is at least one order of magnitude larger than $\sigma_{\sqrt{s}=m_{Z'}}/m_{Z'}$ except for couplings $g_{\mu}$ as small as $0.08$~\footnote{Note that such small coupling would be excluded by relic density and dSph constraints if the DM decay channel is open. For the four benchmarks in Table~\ref{tab1}, this is easily observed in Fig.~\ref{resultados}.}. Therefore, it is reasonable to use the approximation $\Gamma_{Z'}\gg \sigma_{\sqrt{s}=m_{Z'}}$ to compute the resonant effective cross section at $\sqrt{s}=m_{Z'}$. In addition, in the regime $\Gamma_{Z'}\gg \sigma_{\sqrt{s}=m_{Z'}}$ the effective cross section turns out to be independent of the resolution $R$. The corresponding expression is given by,
\begin{equation}
\label{xseff}
\bar{\sigma}_{Z'}=\frac{12\pi\,m^2_{Z'}}{(m^2_{Z'}-(m_C+m_D)^2)(m^2_{Z'}-(m_C-m_D)^2)}\,\mathrm{BR}(Z'\to\mu^+\mu^-)\mathrm{BR}(Z'\to CD),
\end{equation}
where $CD=Z\gamma$ or $ZZ$, with $m_C=m_Z,m_D=0$ or $m_C=m_D=m_Z$, respectively.

From Eq.~(\ref{xseff}) we see that the effective cross section depends on the branching ratios of $Z'\to\mu^+\mu^-$ and $Z'\to Z\gamma/ZZ$. The former can be approximated by $1-\mathrm{BR}_{\mathrm{inv}}$, with $\mathrm{BR}_{\mathrm{inv}}$ given in Eq.~(\ref{eq5}), due to the smallness of the anomalous decays, while the latter can be obtained by computing the partial width corresponding to the anomalous decay channel  and approximating the total decay width by Eq.~(\ref{eq6}). Therefore, in contrast to the processes studied in the previous section for the hadron collider, in this case we do not need to rely on the Goldstone equivalence theorem to estimate the cross sections.

The decay rate for the process $Z'\to Z\gamma$ is
\begin{equation}
\label{WidthZg}
\Gamma(Z'\to Z\gamma)=\frac{p}{32\pi^2m^2_{Z'}}\int \overline{|\mathcal{M}_{Z'Z\gamma}|^2}\,d\Omega,
\end{equation}
where $p=(m^2_{Z'}-m^2_{Z})/2m_{Z'}$ is the momentum of the final state vectors in the center-of-mass frame and $\overline{|\mathcal{M}|^2}$ is the spin-averaged squared of the matrix element which, following the procedure described in~\cite{Racioppi:2009yxa,Anastasopoulos:2008jt,Anastasopoulos:2006cz,Kiritsis:2002aj}, is given by

\begin{equation}
\label{AZg}
\overline{|\mathcal{M}_{Z'Z\gamma}|^2}=g^{\prime 2}\,g^2_Z\,e^2\, \frac{(m^2_{Z'}-m^2_{Z})^2(m^2_{Z'}+m^2_{Z})}{96\pi^4m^2_{Z'}m^2_{Z}}\left[t^{Z'Z\gamma}_{\mu}\, (I_{3\mu}+I_{5\mu})\,m^2_{Z}\right]^2,
\end{equation}
where $e= g_1g_2/\sqrt{g_1^2+g_2^2}$ and $g_Z=\sqrt{g_1^2+g_2^2}$, with $g_1$ and $g_2$ the SM gauge coupling constants of $U(1)$ and $SU(2)$, respectively, and $t^{Z'Z\gamma}_{\mu}= (1-4\sin^2\theta_W)Q_{\mu}$. The expressions of the integrals $I_{i\mu}$ can be read from the general formulae provided in the Appendix~\ref{appendix:a}.

For the process $Z'\to ZZ$ the momentum of the outgoing $Z$ bosons is $p= \sqrt{m^2_{Z'}-4m^2_{Z}}/2$ and the spin-averaged squared of the matrix element is in this case given by~\cite{Racioppi:2009yxa,Anastasopoulos:2008jt},
\begin{equation}
\label{AZZ}
\overline{|\mathcal{M}_{Z'ZZ}|^2}=g^{\prime 2}\,g^4_Z\,\frac{(m^2_{Z'}-4m^2_{Z})^2}{192\pi^4m^2_{Z}}\left[\left(t^{Z'ZZ}_{\mu}(I_{3\mu}+I_{5\mu})+t^{Z'ZZ}_{\nu_{\mu}}(I_{3\nu_{\mu}}+I_{5\nu_{\mu}})\right)\,m^2_{Z}\right]^2,
\end{equation}
where
\begin{eqnarray}
t^{Z'ZZ}_{\mu} & = & \left(2\left(-\frac{1}{2}+2\sin^2\theta_W\right)^2+\frac{1}{2}\right)Q_{\mu}, \\
t^{Z'ZZ}_{\nu_{\mu}} & = & Q_{\mu},
\end{eqnarray}
and we have neglected terms that are suppressed by $(m_\mu/m_Z)^2$.
Again, the expressions of the integrals appearing in Eq.~(\ref{AZZ}) can be easily derived from the formulae in the Appendix~\ref{appendix:a}.

\begin{figure}[t]
    \centering
    \includegraphics[scale=0.4]{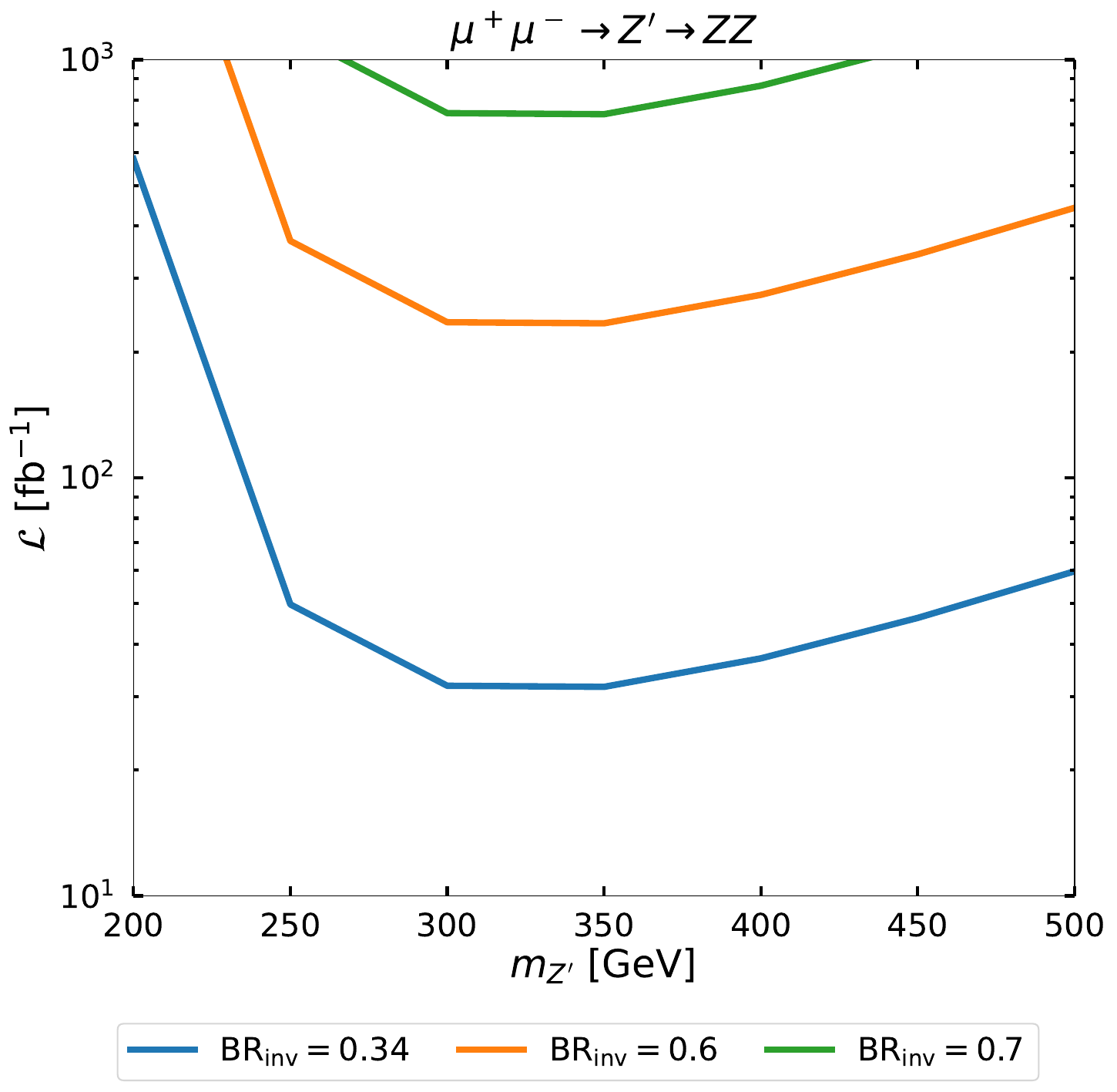}
    \caption{Luminosity required to reach a signal significance of $5\sigma$ for three different values of $\mathrm{BR}_{\mathrm{inv}}$. }
    \label{fig:lumi}
\end{figure}

By computing the effective cross sections of the $ZZ$ and $Z\gamma$ channel from Eq.~(\ref{xseff}), we found that the former is between one and two orders of magnitude larger within the range of $Z'$ masses considered here. For example, with ${\rm BR_{inv}}=0.34$ we obtain cross sections in the range of $(7.5-25.8)\;{\rm fb}$ for the $ZZ$ channel, while for the $Z\gamma$ channel these values are within $(0.07-0.59)\;{\rm fb}$.
In addition,the cross section we obtain for the irreducible $ZZ$ background is in the range of $(0.41-1.31)\;{\rm pb}$ for the $\sqrt{s}$ values considered, which is significantly lower than the $Z\gamma$ cross section values of $(10.6-79.0)\;{\rm pb}$. 
Then, the detection of the $Z\gamma$ channel appears to be very challenging in comparison with the $ZZ$ channel. For this reason in the following we concentrate only in this channel. In order to provide an estimation of the discovery prospects we compute the luminosity required for the significance $s/\sqrt{b}$ to be 5, with the number of background events obtained from the cross section of the irreducible background simulated with \texttt{MadGraph} \footnote{Note that for this estimation we are considering both the signal and the irreducible background inclusively. For a discussion of the potential final states arising from the ZZ decay see for example the case b) in the Appendix B of Ref.~\cite{Barger:1996jm}.}. The results are shown in Fig.~\ref{fig:lumi} for $\mathrm{BR}_{\mathrm{inv}}$ of 0.34, 0.6 and 0.7. It is clear that the required luminosity increases with $\mathrm{BR}_{\mathrm{inv}}$. This behavior is expected: on the one hand, the effective cross section is proportional to $\mathrm{BR}(Z'\to \mu^+\mu^-)= 1-\mathrm{BR}_{\mathrm{inv}}$, and on top of this the total width of the $Z'$ used to compute $\mathrm{BR}(Z'\to ZZ)$ increases with $\mathrm{BR}_{\mathrm{inv}}$ as can be seen by rewritting Eq.~(\ref{eq6}) as $\Gamma_{Z'}= g^2_{\mu}m_{Z'}/(12\pi(1-\mathrm{BR}_{\mathrm{inv}}))$. Therefore, the most promising scenario corresponds to a invisible decay rate dominated by the neutrino final state. 

For each value of $\mathrm{BR}_{\mathrm{inv}}$, the minimum required luminosity is reached at $m_{Z'}=300$ GeV which is consistent with the fact that the anomalous partial decay width $\Gamma(Z'\to ZZ)$ exhibits a peak around that mass. For the most promising scenario, luminosities between $\sim 30\,\mathrm{fb}^{-1}$ and $\sim 60\,\mathrm{fb}^{-1}$ would be enough to detect the anomalous decay channel for $Z'$ masses above 250 GeV. Assuming a total integrated luminosity of $20\,\mathrm{fb}^{-1}$ per year~\cite{Barger:1996jm}, this would correspond to 1.5 $-$ 3 years of data taking. For scenarios in which the DM channel contributes significantly to the invisible decay rate the situation worsens dramatically, with the required luminosity being pushed to values above $200\,\mathrm{fb}^{-1}$ ($\mathrm{BR}_{\mathrm{inv}}=0.6$) or even close to $1\,\mathrm{ab}^{-1}$ ($\mathrm{BR}_{\mathrm{inv}}=0.7$). It is important to emphasize that these are preliminary results and that the developement of a dedicated search strategy with a thorough treatment of the backgrounds is needed to obtain conclusive results. We leave the implementation of such search strategy for future work.


\section{Conclusions}
\label{conclu}
Guided by the current null findings of DM signals, beyond its gravitational influence, which exert strong constraints on WIMP DM models, we embarked on the construction of a DM theory in which second generation leptons and the DM are suitably charged under a new spontaneously broken Abelian $U(1)'_{\mu}$  gauge group such that  the WIMP paradigm is attainable and at the same time all current direct detection, indirect detection and collider experimental constraints are satisfied. In fact, by construction due to the axial nature of the interaction between the DM and the muon, the theory predicts such small contributions to spin independent DM-nuclei scattering cross sections that they are buried under what is known as the neutrino floor. The strongest current constraints on the model are indirectly related to DM and, in those regions of parameter space where the relic density does not overclose the Universe, they come from contributions beyond the SM to the muon anomalous magnetic moment $(g-2)_{\mu}$ and to neutrino trident production.

We studied all these experimental constraints, focusing in particular on the current searches done at the LHC involving leptons and missing energy and showed that there remain large regions of parameter space that evade all of them and for which the DM content can satisfy the latest DM relic density measurements. We also showed that parts of these unconstrained regions could be probed by future collider measurements at the $\sqrt{s}=14$ TeV LHC for luminosities of 300 $\mathrm{fb}^{-1}$ and 3000 $\mathrm{fb}^{-1}$ (even reaching a discovery level in some regions in this case) with a muon-specific search strategy using an invariant mass window around the $Z'$ mass in the more sensitive $3\mu + E_T^{\mathrm{miss}}$ channel, though some unconstrained regions would remain leading still to an elusive WIMP.

Another very interesting feature of the model is that due to the new abelian charges of the SM second generation leptons, crucial to suppress the spin-independent direct detection cross sections, the model is gauge anomalous and can only be interpreted as a low energy effective theory of a non-anomalous UV theory in which part of the fermion spectrum responsible for anomaly cancellation has been integrated out. An implication of this anomalous nature are triple gauge couplings in the low energy effective theory between the anomalous $Z'$ and the EW gauge bosons of the SM, which due to their loop-nature tend to be suppressed, but if able to be probed lead to a window into the UV physics responsible for the anomaly cancellation. We showed that attempts to probe these anomalous couplings at the LHC even at very high luminosities $\mathcal{L}=3000\;\mathrm{fb}^{-1}$ are extremely challenging. However, should a muon collider be built in the future, we have demonstrated that due to the large on-shell $Z'$ production cross section it would be feasible with relatively low luminosities to probe the anomalous couplings, in particular in the $\mu^{+}\mu^{-}\to Z' \to ZZ$ resonant search. Specifically, for this signal we showed that in scenarios where the invisible branching ratio of the $Z'$ is dominated by the decay into neutrinos, luminosities in the range $(30-60)\,\mathrm{fb}^{-1}$ would be enough to detect the anomalous decay channel, provided that the $Z'$ mass is between 250 and 500 GeV.


\numberwithin{equation}{section}
\appendix
\section{Expressions of loop integrals appearing in anomalous triple gauge couplings}
\label{appendix:a}
Here we provide general expressions of loop integrals necessary to compute the decay rates of $Z'$ into $Z\gamma$ and $ZZ$~\cite{Racioppi:2009yxa}:
\begin{eqnarray}
\label{I3}
I_3(p,q;m_f)&=&-\int^1_0 dx \int^{1-x}_0 dy \,\frac{xy}{y(1-y)p^2+x(1-x)q^2+2xy\, p\cdot q -m^2_f}\\
\label{I5}
I_5(p,q;m_f)&=&-\int^1_0 dx \int^{1-x}_0 dy \,\frac{y(y-1)}{y(1-y)p^2+x(1-x)q^2+2xy p\cdot q -m^2_f}\\
\label{I0}
I_0(p,q;m_f)&=&-\int^1_0 dx \int^{1-x}_0 dy \,\frac{1}{y(1-y)p^2+x(1-x)q^2+2xy\, p\cdot q -m^2_f},
\end{eqnarray}
where $p$ and $q$ are the four momenta of the outgoing gauge bosons and $f$ is the fermion running in the loop.
The integrals appearing in Eqs.~(\ref{AZg})-(\ref{AZZ}) are easily obtained from Eqs.~(\ref{I3})-(\ref{I0}). For example, the expression for $I_{3\mu}$ in Eq.~(\ref{AZg}) can be read from Eq.~(\ref{I3}) through the replacements $m_f=m_{\mu},\, p^2=m^2_Z,\, q^2=0$ and $p\cdot q= (m^2_{Z'}-m^2_Z)/2$. Similarly, the integral $I_{3\mu}$ in the case of the decay rate into $ZZ$ corresponds to Eq.~(\ref{I3}) with the replacements $m_f=m_{\mu}, \,p^2=q^2=m^2_Z$ and $p\cdot q= (m^2_{Z'}-2\,m^2_Z)/2$.


\section{Example of cutflows for the LHC analysis}\label{appendix:b}

By way of illustration, we provide some cutflows in Tables~\ref{cutflow1} and~\ref{cutflow2} for the dedicated search of the $3\mu+E_T^\mathrm{miss}$ signal described in Section~\ref{sec:collider} and whose results are presented in Section~\ref{Results}. The total integrated luminosity is set at $300\;\mathrm{fb}^{-1}$. Selection cuts include points 1 and 2 of the search strategy in Section~\ref{sec:collider}.

\begin{table}[!h]
	\centering
	\begin{tabular}{lcc}
	\hline\hline
	   & Signal & Background \\ \hline
	Expected & $11.9$ & $12897$ \\ \hline
	Selection cuts & $6.64$ & $1470$ \\
	$E_T^{\rm miss}>100$ GeV & $5.11$ & $268.9$ \\
	$m_{\mu^+\mu^-}>12$ GeV & $5.11$ & $268.4$\\
	$|m_{\mu^+\mu^-}-m_Z|>10$ GeV & $4.99$ & $36.1$ \\
	$m_T>110$ GeV & $4.49$ & $14.6$ \\
	$|m_{\mu^+\mu^-}-m_{Z'}|<0.1\;m_{Z'}$ & $4.26$ & $1.52$ \\ \hline\hline
	\end{tabular}
	\caption{Cutflow of signal events generated for a particular point of BMII, with $m_{Z'}=350$ GeV and $g_\mu=0.46$. The number of generated events is rescaled here to a total integrated luminosity of $300\;\mathrm{fb}^{-1}$. The selection cuts are constituted by: $p^\mu_T>50$~GeV, $|\eta_\mu|<2.4$, $p^j_T>20$~GeV, $|\eta_j|<2.4$, $\Delta R_{j\mu}>0.4$, $N_b=0$, $N_\mu=3$ with net charge $=\pm 1$. The $m_T>110$ GeV cut is applied to the muon that is not included in the pair which better reconstructs the $Z'$ mass. For more details regarding the cut definitions see Section~\ref{sec:collider}.}
	\label{cutflow1}
\end{table}

\begin{table}[!h]
	\centering
	\begin{tabular}{lcc}
	\hline\hline
	   & Signal & Background \\ \hline 
	Expected &  $14.5$ & $12897$ \\ \hline
	Selection cuts   & $8.05$ & $1470$ \\
	$E_T^{\rm miss}>100$ GeV & $4.93$ & $268.9$ \\
	$m_{\mu^+\mu^-}>12$ GeV & $4.93$ & $268.4$\\
	$|m_{\mu^+\mu^-}-m_Z|>10$ GeV & $4.63$ & $36.1$ \\
	$m_T>110$ GeV & $4.15$ & $14.6$ \\
	$|m_{\mu^+\mu^-}-m_{Z'}|<0.1\;m_{Z'}$ & $3.75$ & $2.30$ \\ \hline\hline
	\end{tabular}
	\caption{Cutflow of signal events generated for a particular point of BMIII, with $m_{Z'}=200$ GeV and $g_\mu=0.35$. The number of generated events is rescaled here to a total integrated luminosity of $300\;\mathrm{fb}^{-1}$. The selection cuts are constituted by: $p^\mu_T>50$~GeV, $|\eta_\mu|<2.4$, $p^j_T>20$~GeV, $|\eta_j|<2.4$, $\Delta R_{j\mu}>0.4$, $N_b=0$, $N_\mu=3$ with net charge $=\pm 1$. The $m_T>110$ GeV cut is applied to the muon that is not included in the pair which better reconstructs the $Z'$ mass. For more details regarding the cut definitions see Section~\ref{sec:collider}.}
	\label{cutflow2}
\end{table}

\pagebreak
\section*{Acknowledgments}
The authors would like to thank Mar\'{\i}a Teresa Dova, Carlos Wagner and Hern\'an Wahlberg for useful discussions. This work is  partially supported by CONICET and ANPCyT under projects PICT 2016-0164, PICT 2017-2751, and PICT 2018-03682.

\bibliographystyle{JHEP}
\bibliography{lit2}

\providecommand{\href}[2]{#2}\begingroup\raggedright\begin{thebibliography}{10}

\bibitem{DiGangi:2018dek}
{\scshape XENON} collaboration, \emph{{First dark matter search results of the
  XENON1T experiment}},
  \href{https://doi.org/10.1393/ncc/i2018-18109-5}{\emph{Nuovo Cim. C}
  {\bfseries 41} (2018) 109}.

\bibitem{XENON100:2013ele}
{\scshape XENON100} collaboration, \emph{{Limits on spin-dependent WIMP-nucleon
  cross sections from 225 live days of XENON100 data}},
  \href{https://doi.org/10.1103/PhysRevLett.111.021301}{\emph{Phys. Rev. Lett.}
  {\bfseries 111} (2013) 021301}
  [\href{https://arxiv.org/abs/1301.6620}{{\ttfamily 1301.6620}}].

\bibitem{LUX:2016ggv}
{\scshape LUX} collaboration, \emph{{Results from a search for dark matter in
  the complete LUX exposure}},
  \href{https://doi.org/10.1103/PhysRevLett.118.021303}{\emph{Phys. Rev. Lett.}
  {\bfseries 118} (2017) 021303}
  [\href{https://arxiv.org/abs/1608.07648}{{\ttfamily 1608.07648}}].

\bibitem{Arcadi:2017kky}
G.~Arcadi, M.~Dutra, P.~Ghosh, M.~Lindner, Y.~Mambrini, M.~Pierre et~al.,
  \emph{{The waning of the WIMP? A review of models, searches, and
  constraints}},
  \href{https://doi.org/10.1140/epjc/s10052-018-5662-y}{\emph{Eur. Phys. J. C}
  {\bfseries 78} (2018) 203}
  [\href{https://arxiv.org/abs/1703.07364}{{\ttfamily 1703.07364}}].

\bibitem{GAMBIT:2021rlp}
{\scshape GAMBIT} collaboration, \emph{{Thermal WIMPs and the scale of new
  physics: global fits of Dirac dark matter effective field theories}},
  \href{https://doi.org/10.1140/epjc/s10052-021-09712-6}{\emph{Eur. Phys. J. C}
  {\bfseries 81} (2021) 992}
  [\href{https://arxiv.org/abs/2106.02056}{{\ttfamily 2106.02056}}].

\bibitem{Carena:2019pwq}
M.~Carena, J.~Osborne, N.~R. Shah and C.~E.~M. Wagner, \emph{{Return of the
  WIMP: Missing energy signals and the Galactic Center excess}},
  \href{https://doi.org/10.1103/PhysRevD.100.055002}{\emph{Phys. Rev. D}
  {\bfseries 100} (2019) 055002}
  [\href{https://arxiv.org/abs/1905.03768}{{\ttfamily 1905.03768}}].

\bibitem{Blanco:2019hah}
C.~Blanco, M.~Escudero, D.~Hooper and S.~J. Witte, \emph{{Z' mediated WIMPs:
  dead, dying, or soon to be detected?}},
  \href{https://doi.org/10.1088/1475-7516/2019/11/024}{\emph{JCAP} {\bfseries
  11} (2019) 024} [\href{https://arxiv.org/abs/1907.05893}{{\ttfamily
  1907.05893}}].

\bibitem{Fermi-LAT:2015sau}
{\scshape Fermi-LAT} collaboration, \emph{{Fermi-LAT Observations of
  High-Energy $\gamma$-Ray Emission Toward the Galactic Center}},
  \href{https://doi.org/10.3847/0004-637X/819/1/44}{\emph{Astrophys. J.}
  {\bfseries 819} (2016) 44}
  [\href{https://arxiv.org/abs/1511.02938}{{\ttfamily 1511.02938}}].

\bibitem{Fermi-LAT:2013sme}
{\scshape Fermi-LAT} collaboration, \emph{{Dark Matter Constraints from
  Observations of 25 Milky Way Satellite Galaxies with the Fermi Large Area
  Telescope}}, \href{https://doi.org/10.1103/PhysRevD.89.042001}{\emph{Phys.
  Rev. D} {\bfseries 89} (2014) 042001}
  [\href{https://arxiv.org/abs/1310.0828}{{\ttfamily 1310.0828}}].

\bibitem{Fermi-LAT:2015att}
{\scshape Fermi-LAT} collaboration, \emph{{Searching for Dark Matter
  Annihilation from Milky Way Dwarf Spheroidal Galaxies with Six Years of Fermi
  Large Area Telescope Data}},
  \href{https://doi.org/10.1103/PhysRevLett.115.231301}{\emph{Phys. Rev. Lett.}
  {\bfseries 115} (2015) 231301}
  [\href{https://arxiv.org/abs/1503.02641}{{\ttfamily 1503.02641}}].

\bibitem{ATLAS:2016bek}
{\scshape ATLAS} collaboration, \emph{{Search for new phenomena in final states
  with an energetic jet and large missing transverse momentum in $pp$
  collisions at $\sqrt{s}=13$ TeV using the ATLAS detector}},
  \href{https://doi.org/10.1103/PhysRevD.94.032005}{\emph{Phys. Rev. D}
  {\bfseries 94} (2016) 032005}
  [\href{https://arxiv.org/abs/1604.07773}{{\ttfamily 1604.07773}}].

\bibitem{CMS:2017jdm}
{\scshape CMS} collaboration, \emph{{Search for dark matter produced with an
  energetic jet or a hadronically decaying W or Z boson at $ \sqrt{s}=13 $
  TeV}}, \href{https://doi.org/10.1007/JHEP07(2017)014}{\emph{JHEP} {\bfseries
  07} (2017) 014} [\href{https://arxiv.org/abs/1703.01651}{{\ttfamily
  1703.01651}}].

\bibitem{ATLAS:2017txd}
{\scshape ATLAS} collaboration, \emph{{Measurement of detector-corrected
  observables sensitive to the anomalous production of events with jets and
  large missing transverse momentum in $pp$ collisions at $\mathbf
  {\sqrt{s}=13}$ TeV using the ATLAS detector}},
  \href{https://doi.org/10.1140/epjc/s10052-017-5315-6}{\emph{Eur. Phys. J. C}
  {\bfseries 77} (2017) 765}
  [\href{https://arxiv.org/abs/1707.03263}{{\ttfamily 1707.03263}}].

\bibitem{ATLAS:2017bfj}
{\scshape ATLAS} collaboration, \emph{{Search for dark matter and other new
  phenomena in events with an energetic jet and large missing transverse
  momentum using the ATLAS detector}},
  \href{https://doi.org/10.1007/JHEP01(2018)126}{\emph{JHEP} {\bfseries 01}
  (2018) 126} [\href{https://arxiv.org/abs/1711.03301}{{\ttfamily
  1711.03301}}].

\bibitem{CMS:2014jvv}
{\scshape CMS} collaboration, \emph{{Search for dark matter, extra dimensions,
  and unparticles in monojet events in proton\textendash{}proton collisions at
  $\sqrt{s} = 8$ TeV}},
  \href{https://doi.org/10.1140/epjc/s10052-015-3451-4}{\emph{Eur. Phys. J. C}
  {\bfseries 75} (2015) 235} [\href{https://arxiv.org/abs/1408.3583}{{\ttfamily
  1408.3583}}].

\bibitem{Fox:2008kb}
P.~J. Fox and E.~Poppitz, \emph{{Leptophilic Dark Matter}},
  \href{https://doi.org/10.1103/PhysRevD.79.083528}{\emph{Phys. Rev. D}
  {\bfseries 79} (2009) 083528}
  [\href{https://arxiv.org/abs/0811.0399}{{\ttfamily 0811.0399}}].

\bibitem{Haisch:2013uaa}
U.~Haisch and F.~Kahlhoefer, \emph{{On the importance of loop-induced
  spin-independent interactions for dark matter direct detection}},
  \href{https://doi.org/10.1088/1475-7516/2013/04/050}{\emph{JCAP} {\bfseries
  04} (2013) 050} [\href{https://arxiv.org/abs/1302.4454}{{\ttfamily
  1302.4454}}].

\bibitem{Kopp:2009et}
J.~Kopp, V.~Niro, T.~Schwetz and J.~Zupan, \emph{{DAMA/LIBRA and leptonically
  interacting Dark Matter}},
  \href{https://doi.org/10.1103/PhysRevD.80.083502}{\emph{Phys. Rev. D}
  {\bfseries 80} (2009) 083502}
  [\href{https://arxiv.org/abs/0907.3159}{{\ttfamily 0907.3159}}].

\bibitem{DEramo:2014nmf}
F.~D'Eramo and M.~Procura, \emph{{Connecting Dark Matter UV Complete Models to
  Direct Detection Rates via Effective Field Theory}},
  \href{https://doi.org/10.1007/JHEP04(2015)054}{\emph{JHEP} {\bfseries 04}
  (2015) 054} [\href{https://arxiv.org/abs/1411.3342}{{\ttfamily 1411.3342}}].

\bibitem{DEramo:2017zqw}
F.~D'Eramo, B.~J. Kavanagh and P.~Panci, \emph{{Probing Leptophilic Dark
  Sectors with Hadronic Processes}},
  \href{https://doi.org/10.1016/j.physletb.2017.05.063}{\emph{Phys. Lett. B}
  {\bfseries 771} (2017) 339}
  [\href{https://arxiv.org/abs/1702.00016}{{\ttfamily 1702.00016}}].

\bibitem{Dror:2017nsg}
J.~A. Dror, R.~Lasenby and M.~Pospelov, \emph{{Dark forces coupled to
  nonconserved currents}},
  \href{https://doi.org/10.1103/PhysRevD.96.075036}{\emph{Phys. Rev. D}
  {\bfseries 96} (2017) 075036}
  [\href{https://arxiv.org/abs/1707.01503}{{\ttfamily 1707.01503}}].

\bibitem{Ismail:2017ulg}
A.~Ismail, A.~Katz and D.~Racco, \emph{{On dark matter interactions with the
  Standard Model through an anomalous $Z'$}},
  \href{https://doi.org/10.1007/JHEP10(2017)165}{\emph{JHEP} {\bfseries 10}
  (2017) 165} [\href{https://arxiv.org/abs/1707.00709}{{\ttfamily
  1707.00709}}].

\bibitem{Ismail:2017fgq}
A.~Ismail and A.~Katz, \emph{{Anomalous $Z'$ and diboson resonances at the
  LHC}}, \href{https://doi.org/10.1007/JHEP04(2018)122}{\emph{JHEP} {\bfseries
  04} (2018) 122} [\href{https://arxiv.org/abs/1712.01840}{{\ttfamily
  1712.01840}}].

\bibitem{Kawamura:2019rth}
J.~Kawamura, S.~Raby and A.~Trautner, \emph{{Complete vectorlike fourth family
  and new U(1)' for muon anomalies}},
  \href{https://doi.org/10.1103/PhysRevD.100.055030}{\emph{Phys. Rev. D}
  {\bfseries 100} (2019) 055030}
  [\href{https://arxiv.org/abs/1906.11297}{{\ttfamily 1906.11297}}].

\bibitem{Abdughani:2021oit}
M.~Abdughani, Y.-Z. Fan, C.-T. Lu, T.-P. Tang and Y.-L.~S. Tsai,
  \emph{{Muonphilic Dark Matter explanation of gamma-ray galactic center
  excess: a comprehensive analysis}},
  \href{https://arxiv.org/abs/2111.02946}{{\ttfamily 2111.02946}}.

\bibitem{Perelstein:2020suc}
M.~Perelstein and Y.~C. San, \emph{{Dark Matter as a Solution to Muonic
  Puzzles}}, \href{https://doi.org/10.1103/PhysRevD.103.035032}{\emph{Phys.
  Rev. D} {\bfseries 103} (2021) 035032}
  [\href{https://arxiv.org/abs/2009.09867}{{\ttfamily 2009.09867}}].

\bibitem{Planck:2018vyg}
{\scshape Planck} collaboration, \emph{{Planck 2018 results. VI. Cosmological
  parameters}},
  \href{https://doi.org/10.1051/0004-6361/201833910}{\emph{Astron. Astrophys.}
  {\bfseries 641} (2020) A6}
  [\href{https://arxiv.org/abs/1807.06209}{{\ttfamily 1807.06209}}].

\bibitem{Belanger:2010pz}
G.~Belanger, F.~Boudjema, A.~Pukhov and A.~Semenov, \emph{{micrOMEGAs: A Tool
  for dark matter studies}},
  \href{https://doi.org/10.1393/ncc/i2010-10591-3}{\emph{Nuovo Cim. C}
  {\bfseries 033N2} (2010) 111}
  [\href{https://arxiv.org/abs/1005.4133}{{\ttfamily 1005.4133}}].

\bibitem{ParticleDataGroup:2020ssz}
{\scshape Particle Data Group} collaboration, \emph{{Review of Particle
  Physics}}, \href{https://doi.org/10.1093/ptep/ptaa104}{\emph{PTEP} {\bfseries
  2020} (2020) 083C01}.

\bibitem{Lin:2019uvt}
T.~Lin, \emph{{Dark matter models and direct detection}},
  \href{https://doi.org/10.22323/1.333.0009}{\emph{PoS} {\bfseries 333} (2019)
  009} [\href{https://arxiv.org/abs/1904.07915}{{\ttfamily 1904.07915}}].

\bibitem{Aprile_2018}
{\scshape XENON} collaboration, \emph{{Dark Matter Search Results from a One
  Ton-Year Exposure of XENON1T}},
  \href{https://doi.org/10.1103/PhysRevLett.121.111302}{\emph{Phys. Rev. Lett.}
  {\bfseries 121} (2018) 111302}
  [\href{https://arxiv.org/abs/1805.12562}{{\ttfamily 1805.12562}}].

\bibitem{Mayet:2016zxu}
F.~Mayet et~al., \emph{{A review of the discovery reach of directional Dark
  Matter detection}},
  \href{https://doi.org/10.1016/j.physrep.2016.02.007}{\emph{Phys. Rept.}
  {\bfseries 627} (2016) 1} [\href{https://arxiv.org/abs/1602.03781}{{\ttfamily
  1602.03781}}].

\bibitem{2020}
A.~Mohamadnejad, \emph{{Gravitational waves from scale-invariant vector dark
  matter model: Probing below the neutrino-floor}},
  \href{https://doi.org/10.1140/epjc/s10052-020-7756-6}{\emph{Eur. Phys. J. C}
  {\bfseries 80} (2020) 197}
  [\href{https://arxiv.org/abs/1907.08899}{{\ttfamily 1907.08899}}].

\bibitem{Wu:2017iji}
Y.~Wu, T.~Ma, B.~Zhang and G.~Cacciapaglia, \emph{{Composite Dark Matter and
  Higgs}}, \href{https://doi.org/10.1007/JHEP11(2017)058}{\emph{JHEP}
  {\bfseries 11} (2017) 058}
  [\href{https://arxiv.org/abs/1703.06903}{{\ttfamily 1703.06903}}].

\bibitem{Hoof_2020}
S.~Hoof, A.~Geringer-Sameth and R.~Trotta, \emph{A global analysis of dark
  matter signals from 27 dwarf spheroidal galaxies using 11 years of fermi-lat
  observations},
  \href{https://doi.org/10.1088/1475-7516/2020/02/012}{\emph{Journal of
  Cosmology and Astroparticle Physics} {\bfseries 2020} (2020) 012–012}.

\bibitem{PAMELA:2010kea}
{\scshape PAMELA} collaboration, \emph{{PAMELA results on the cosmic-ray
  antiproton flux from 60 MeV to 180 GeV in kinetic energy}},
  \href{https://doi.org/10.1103/PhysRevLett.105.121101}{\emph{Phys. Rev. Lett.}
  {\bfseries 105} (2010) 121101}
  [\href{https://arxiv.org/abs/1007.0821}{{\ttfamily 1007.0821}}].

\bibitem{AMS:2021nhj}
{\scshape AMS} collaboration, \emph{{The Alpha Magnetic Spectrometer (AMS) on
  the international space station: Part II \textemdash{} Results from the first
  seven years}},
  \href{https://doi.org/10.1016/j.physrep.2020.09.003}{\emph{Phys. Rept.}
  {\bfseries 894} (2021) 1}.

\bibitem{Bergstr_m_2013}
L.~Bergstrom, T.~Bringmann, I.~Cholis, D.~Hooper and C.~Weniger, \emph{{New
  Limits on Dark Matter Annihilation from AMS Cosmic Ray Positron Data}},
  \href{https://doi.org/10.1103/PhysRevLett.111.171101}{\emph{Phys. Rev. Lett.}
  {\bfseries 111} (2013) 171101}
  [\href{https://arxiv.org/abs/1306.3983}{{\ttfamily 1306.3983}}].

\bibitem{john2021cosmicray}
I.~John and T.~Linden, \emph{{Cosmic-ray positrons strongly constrain
  leptophilic dark matter}},
  \href{https://doi.org/10.1088/1475-7516/2021/12/007}{\emph{JCAP} {\bfseries
  12} (2021) 007} [\href{https://arxiv.org/abs/2107.10261}{{\ttfamily
  2107.10261}}].

\bibitem{Leane:2018kjk}
R.~K. Leane, T.~R. Slatyer, J.~F. Beacom and K.~C.~Y. Ng, \emph{{GeV-scale
  thermal WIMPs: Not even slightly ruled out}},
  \href{https://doi.org/10.1103/PhysRevD.98.023016}{\emph{Phys. Rev. D}
  {\bfseries 98} (2018) 023016}
  [\href{https://arxiv.org/abs/1805.10305}{{\ttfamily 1805.10305}}].

\bibitem{Muong-2:2006rrc}
{\scshape Muon g-2} collaboration, \emph{{Final Report of the Muon E821
  Anomalous Magnetic Moment Measurement at BNL}},
  \href{https://doi.org/10.1103/PhysRevD.73.072003}{\emph{Phys. Rev. D}
  {\bfseries 73} (2006) 072003}
  [\href{https://arxiv.org/abs/hep-ex/0602035}{{\ttfamily hep-ex/0602035}}].

\bibitem{Muong-2:2021ojo}
{\scshape Muon g-2} collaboration, \emph{{Measurement of the Positive Muon
  Anomalous Magnetic Moment to 0.46 ppm}},
  \href{https://doi.org/10.1103/PhysRevLett.126.141801}{\emph{Phys. Rev. Lett.}
  {\bfseries 126} (2021) 141801}
  [\href{https://arxiv.org/abs/2104.03281}{{\ttfamily 2104.03281}}].

\bibitem{Aoyama:2020ynm}
T.~Aoyama et~al., \emph{{The anomalous magnetic moment of the muon in the
  Standard Model}},
  \href{https://doi.org/10.1016/j.physrep.2020.07.006}{\emph{Phys. Rept.}
  {\bfseries 887} (2020) 1} [\href{https://arxiv.org/abs/2006.04822}{{\ttfamily
  2006.04822}}].

\bibitem{Borsanyi_2021}
S.~Borsanyi, Z.~Fodor, J.~N. Guenther, C.~Hoelbling, S.~D. Katz, L.~Lellouch
  et~al., \emph{Leading hadronic contribution to the muon magnetic moment from
  lattice qcd}, \href{https://doi.org/10.1038/s41586-021-03418-1}{\emph{Nature}
  {\bfseries 593} (2021) 51–55}.

\bibitem{Balkin:2021rvh}
R.~Balkin, C.~Delaunay, M.~Geller, E.~Kajomovitz, G.~Perez, Y.~Shpilman et~al.,
  \emph{{Custodial symmetry for muon g-2}},
  \href{https://doi.org/10.1103/PhysRevD.104.053009}{\emph{Phys. Rev. D}
  {\bfseries 104} (2021) 053009}
  [\href{https://arxiv.org/abs/2104.08289}{{\ttfamily 2104.08289}}].

\bibitem{CCFR:1991lpl}
{\scshape CCFR} collaboration, \emph{{Neutrino tridents and W Z interference}},
  \href{https://doi.org/10.1103/PhysRevLett.66.3117}{\emph{Phys. Rev. Lett.}
  {\bfseries 66} (1991) 3117}.

\bibitem{Altmannshofer:2019zhy}
W.~Altmannshofer, S.~Gori, J.~Mart\'\i{}n-Albo, A.~Sousa and M.~Wallbank,
  \emph{{Neutrino Tridents at DUNE}},
  \href{https://doi.org/10.1103/PhysRevD.100.115029}{\emph{Phys. Rev. D}
  {\bfseries 100} (2019) 115029}
  [\href{https://arxiv.org/abs/1902.06765}{{\ttfamily 1902.06765}}].

\bibitem{Bell:2014tta}
N.~F. Bell, Y.~Cai, R.~K. Leane and A.~D. Medina, \emph{{Leptophilic dark
  matter with $Z'$ interactions}},
  \href{https://doi.org/10.1103/PhysRevD.90.035027}{\emph{Phys. Rev. D}
  {\bfseries 90} (2014) 035027}
  [\href{https://arxiv.org/abs/1407.3001}{{\ttfamily 1407.3001}}].

\bibitem{Ma:2001md}
E.~Ma, D.~P. Roy and S.~Roy, \emph{{Gauged $L_\mu-L_\tau$ with large muon
  anomalous magnetic moment and the bimaximal mixing of neutrinos}},
  \href{https://doi.org/10.1016/S0370-2693(01)01428-9}{\emph{Phys. Lett. B}
  {\bfseries 525} (2002) 101}
  [\href{https://arxiv.org/abs/hep-ph/0110146}{{\ttfamily hep-ph/0110146}}].

\bibitem{delAguila:2014soa}
F.~del Aguila, M.~Chala, J.~Santiago and Y.~Yamamoto, \emph{{Collider limits on
  leptophilic interactions}},
  \href{https://doi.org/10.1007/JHEP03(2015)059}{\emph{JHEP} {\bfseries 03}
  (2015) 059} [\href{https://arxiv.org/abs/1411.7394}{{\ttfamily 1411.7394}}].

\bibitem{Drees:2018hhs}
M.~Drees, M.~Shi and Z.~Zhang, \emph{{Constraints on $U(1)_{L_\mu-L_\tau}$ from
  LHC Data}}, \href{https://doi.org/10.1016/j.physletb.2019.02.029}{\emph{Phys.
  Lett. B} {\bfseries 791} (2019) 130}
  [\href{https://arxiv.org/abs/1811.12446}{{\ttfamily 1811.12446}}].

\bibitem{Conte:2014zja}
E.~Conte, B.~Dumont, B.~Fuks and C.~Wymant, \emph{{Designing and recasting LHC
  analyses with MadAnalysis 5}},
  \href{https://doi.org/10.1140/epjc/s10052-014-3103-0}{\emph{Eur. Phys. J. C}
  {\bfseries 74} (2014) 3103}
  [\href{https://arxiv.org/abs/1405.3982}{{\ttfamily 1405.3982}}].

\bibitem{Dumont:2014tja}
B.~Dumont, B.~Fuks, S.~Kraml, S.~Bein, G.~Chalons, E.~Conte et~al.,
  \emph{{Toward a public analysis database for LHC new physics searches using
  MADANALYSIS 5}},
  \href{https://doi.org/10.1140/epjc/s10052-014-3242-3}{\emph{Eur. Phys. J. C}
  {\bfseries 75} (2015) 56} [\href{https://arxiv.org/abs/1407.3278}{{\ttfamily
  1407.3278}}].

\bibitem{Conte:2018vmg}
E.~Conte and B.~Fuks, \emph{{Confronting new physics theories to LHC data with
  MADANALYSIS 5}}, \href{https://doi.org/10.1142/S0217751X18300272}{\emph{Int.
  J. Mod. Phys. A} {\bfseries 33} (2018) 1830027}
  [\href{https://arxiv.org/abs/1808.00480}{{\ttfamily 1808.00480}}].

\bibitem{Alwall:2014hca}
J.~Alwall, R.~Frederix, S.~Frixione, V.~Hirschi, F.~Maltoni, O.~Mattelaer
  et~al., \emph{{The automated computation of tree-level and next-to-leading
  order differential cross sections, and their matching to parton shower
  simulations}}, \href{https://doi.org/10.1007/JHEP07(2014)079}{\emph{JHEP}
  {\bfseries 07} (2014) 079} [\href{https://arxiv.org/abs/1405.0301}{{\ttfamily
  1405.0301}}].

\bibitem{Alloul:2013bka}
A.~Alloul, N.~D. Christensen, C.~Degrande, C.~Duhr and B.~Fuks,
  \emph{{FeynRules 2.0 - A complete toolbox for tree-level phenomenology}},
  \href{https://doi.org/10.1016/j.cpc.2014.04.012}{\emph{Comput. Phys. Commun.}
  {\bfseries 185} (2014) 2250}
  [\href{https://arxiv.org/abs/1310.1921}{{\ttfamily 1310.1921}}].

\bibitem{Sjostrand:2014zea}
T.~Sj\"ostrand, S.~Ask, J.~R. Christiansen, R.~Corke, N.~Desai, P.~Ilten
  et~al., \emph{{An introduction to PYTHIA 8.2}},
  \href{https://doi.org/10.1016/j.cpc.2015.01.024}{\emph{Comput. Phys. Commun.}
  {\bfseries 191} (2015) 159}
  [\href{https://arxiv.org/abs/1410.3012}{{\ttfamily 1410.3012}}].

\bibitem{deFavereau:2013fsa}
{\scshape DELPHES 3} collaboration, \emph{{DELPHES 3, A modular framework for
  fast simulation of a generic collider experiment}},
  \href{https://doi.org/10.1007/JHEP02(2014)057}{\emph{JHEP} {\bfseries 02}
  (2014) 057} [\href{https://arxiv.org/abs/1307.6346}{{\ttfamily 1307.6346}}].

\bibitem{Araz:2020dlf}
J.~Y. Araz and B.~Fuks, \emph{{Implementation of the ATLAS-SUSY-2018-32
  analysis (sleptons and electroweakinos with two leptons and missing
  transverse energy; 139 fb$^{-1}$)}},
  \href{https://doi.org/10.1142/S0217732321410054}{\emph{Mod. Phys. Lett. A}
  {\bfseries 36} (2021) 2141005}.

\bibitem{ATLAS:2019lff}
{\scshape ATLAS} collaboration, \emph{{Search for electroweak production of
  charginos and sleptons decaying into final states with two leptons and
  missing transverse momentum in $\sqrt{s}=13$ TeV $pp$ collisions using the
  ATLAS detector}},
  \href{https://doi.org/10.1140/epjc/s10052-019-7594-6}{\emph{Eur. Phys. J. C}
  {\bfseries 80} (2020) 123}
  [\href{https://arxiv.org/abs/1908.08215}{{\ttfamily 1908.08215}}].

\bibitem{cms_sus_16_039}
B.~Fuks and S.~Mondal, \emph{{Implementation of a search for electroweakinos in
  the multi-lepton + missing energy channel (35.9 fb-1; 13 TeV;
  CMS-SUS-16-039)}},  2021.
\newblock 10.14428/DVN/E8WN3G.

\bibitem{CMS:2017moi}
{\scshape CMS} collaboration, \emph{{Search for electroweak production of
  charginos and neutralinos in multilepton final states in proton-proton
  collisions at $\sqrt{s}=$ 13 TeV}},
  \href{https://doi.org/10.1007/JHEP03(2018)166}{\emph{JHEP} {\bfseries 03}
  (2018) 166} [\href{https://arxiv.org/abs/1709.05406}{{\ttfamily
  1709.05406}}].

\bibitem{ATLAS:2013rla}
{\scshape ATLAS Collaboration} collaboration, \emph{{Search for direct
  production of charginos and neutralinos in events with three leptons and
  missing transverse momentum in 21$\,$fb$^{-1}$ of pp collisions at
  $\sqrt{s}=8\,$TeV with the ATLAS detector}},  tech. rep., CERN, Geneva, 2013.

\bibitem{Cowan:2010js}
G.~Cowan, K.~Cranmer, E.~Gross and O.~Vitells, \emph{{Asymptotic formulae for
  likelihood-based tests of new physics}},
  \href{https://doi.org/10.1140/epjc/s10052-011-1554-0}{\emph{Eur. Phys. J. C}
  {\bfseries 71} (2011) 1554}
  [\href{https://arxiv.org/abs/1007.1727}{{\ttfamily 1007.1727}}].

\bibitem{PRESKILL1991323}
J.~Preskill, \emph{Gauge anomalies in an effective field theory},
  \href{https://doi.org/https://doi.org/10.1016/0003-4916(91)90046-B}{\emph{Annals
  of Physics} {\bfseries 210} (1991) 323}.

\bibitem{Dedes:2012me}
A.~Dedes and K.~Suxho, \emph{{Heavy Fermion Non-Decoupling Effects in Triple
  Gauge Boson Vertices}},
  \href{https://doi.org/10.1103/PhysRevD.85.095024}{\emph{Phys. Rev. D}
  {\bfseries 85} (2012) 095024}
  [\href{https://arxiv.org/abs/1202.4940}{{\ttfamily 1202.4940}}].

\bibitem{Batra:2005rh}
P.~Batra, B.~A. Dobrescu and D.~Spivak, \emph{{Anomaly-free sets of fermions}},
  \href{https://doi.org/10.1063/1.2222081}{\emph{J. Math. Phys.} {\bfseries 47}
  (2006) 082301} [\href{https://arxiv.org/abs/hep-ph/0510181}{{\ttfamily
  hep-ph/0510181}}].

\bibitem{Bilal:2008qx}
A.~Bilal, \emph{{Lectures on Anomalies}},
  \href{https://arxiv.org/abs/0802.0634}{{\ttfamily 0802.0634}}.

\bibitem{AlAli:2021let}
H.~Al~Ali et~al., \emph{{The Muon Smasher's Guide}},
  \href{https://arxiv.org/abs/2103.14043}{{\ttfamily 2103.14043}}.

\bibitem{Huang:2021nkl}
G.-y. Huang, F.~S. Queiroz and W.~Rodejohann, \emph{{Gauged
  $L^{}_{\mu}{-}L^{}_{\tau}$ at a muon collider}},
  \href{https://doi.org/10.1103/PhysRevD.103.095005}{\emph{Phys. Rev. D}
  {\bfseries 103} (2021) 095005}
  [\href{https://arxiv.org/abs/2101.04956}{{\ttfamily 2101.04956}}].

\bibitem{Yin:2020afe}
W.~Yin and M.~Yamaguchi, \emph{{Muon $g-2$ at multi-TeV muon collider}},
  \href{https://arxiv.org/abs/2012.03928}{{\ttfamily 2012.03928}}.

\bibitem{Barger:1995hr}
V.~D. Barger, M.~S. Berger, J.~F. Gunion and T.~Han, \emph{{s channel Higgs
  boson production at a muon muon collider}},
  \href{https://doi.org/10.1103/PhysRevLett.75.1462}{\emph{Phys. Rev. Lett.}
  {\bfseries 75} (1995) 1462}
  [\href{https://arxiv.org/abs/hep-ph/9504330}{{\ttfamily hep-ph/9504330}}].

\bibitem{Barger:1996jm}
V.~D. Barger, M.~S. Berger, J.~F. Gunion and T.~Han, \emph{{Higgs Boson physics
  in the s channel at $\mu^+ \mu^-$ colliders}},
  \href{https://doi.org/10.1016/S0370-1573(96)00041-5}{\emph{Phys. Rept.}
  {\bfseries 286} (1997) 1}
  [\href{https://arxiv.org/abs/hep-ph/9602415}{{\ttfamily hep-ph/9602415}}].

\bibitem{Han:2021lnp}
T.~Han, W.~Kilian, N.~Kreher, Y.~Ma, J.~Reuter, T.~Striegl et~al.,
  \emph{{Precision Test of the Muon-Higgs Coupling at a High-energy Muon
  Collider}},  \href{https://arxiv.org/abs/2108.05362}{{\ttfamily 2108.05362}}.

\bibitem{Delahaye:2019omf}
J.~P. Delahaye, M.~Diemoz, K.~Long, B.~Mansouli\'e, N.~Pastrone, L.~Rivkin
  et~al., \emph{{Muon Colliders}},
  \href{https://arxiv.org/abs/1901.06150}{{\ttfamily 1901.06150}}.

\bibitem{Bartosik:2020xwr}
N.~Bartosik et~al., \emph{{Detector and Physics Performance at a Muon
  Collider}},
  \href{https://doi.org/10.1088/1748-0221/15/05/P05001}{\emph{JINST} {\bfseries
  15} (2020) P05001} [\href{https://arxiv.org/abs/2001.04431}{{\ttfamily
  2001.04431}}].

\bibitem{Zaza:2021bj}
A.~Zaza, A.~Colaleo, F.~Errico, P.~Mastrapasqua and R.~Venditti, \emph{{Search
  for $H \rightarrow ZZ^{*} \rightarrow 4\mu$ at a Multi-TeV Muon Collider}},
  \href{https://doi.org/10.22323/1.397.0203}{\emph{PoS} {\bfseries LHCP2021}
  (2021) 203}.

\bibitem{Racioppi:2009yxa}
A.~Racioppi, \emph{{Anomalies, U (1)' and the MSSM}}, Ph.D. thesis, Rome U.,Tor
  Vergata, 2009.
\newblock \href{https://arxiv.org/abs/0907.1535}{{\ttfamily 0907.1535}}.

\bibitem{Anastasopoulos:2008jt}
P.~Anastasopoulos, F.~Fucito, A.~Lionetto, G.~Pradisi, A.~Racioppi and Y.~S.
  Stanev, \emph{{Minimal Anomalous U(1)-prime Extension of the MSSM}},
  \href{https://doi.org/10.1103/PhysRevD.78.085014}{\emph{Phys. Rev. D}
  {\bfseries 78} (2008) 085014}
  [\href{https://arxiv.org/abs/0804.1156}{{\ttfamily 0804.1156}}].

\bibitem{Anastasopoulos:2006cz}
P.~Anastasopoulos, M.~Bianchi, E.~Dudas and E.~Kiritsis, \emph{{Anomalies,
  anomalous U(1)'s and generalized Chern-Simons terms}},
  \href{https://doi.org/10.1088/1126-6708/2006/11/057}{\emph{JHEP} {\bfseries
  11} (2006) 057} [\href{https://arxiv.org/abs/hep-th/0605225}{{\ttfamily
  hep-th/0605225}}].

\bibitem{Kiritsis:2002aj}
E.~Kiritsis and P.~Anastasopoulos, \emph{{The Anomalous magnetic moment of the
  muon in the D-brane realization of the standard model}},
  \href{https://doi.org/10.1088/1126-6708/2002/05/054}{\emph{JHEP} {\bfseries
  05} (2002) 054} [\href{https://arxiv.org/abs/hep-ph/0201295}{{\ttfamily
  hep-ph/0201295}}].

\end{thebibliography}\endgroup

\end{document}